\documentclass[11pt]{article}
\usepackage[utf8]{inputenc} 
\usepackage[T1]{fontenc}    
\usepackage{}
\usepackage{url,booktabs,amsfonts,nicefrac,microtype,graphicx,amsmath,bm,xcolor,algorithm2e,booktabs,mathtools,enumitem,mathrsfs,amssymb,soul,subfiles,tikz,lscape,titling} 
\usepackage[symbol]{footmisc}

\usepackage[hidelinks]{hyperref}
\usepackage[margin=1in]{geometry}
\graphicspath{ {./images/} }
\allowdisplaybreaks

\usepackage[backend=biber,style=numeric,citestyle=authoryear,sorting=none,giveninits=true,maxcitenames=1,uniquelist=false,maxbibnames=10,natbib=true]{biblatex}
\bibliography{ref}

\newtheorem{theorem}{Theorem}

\newtheorem{assumption}{Assumption}
\newtheorem{proposition}{Proposition}
\newtheorem{corollary}{Corollary}
\newtheorem{Algorithm}{Algorithm}
\newtheorem{remark}{Remark}

\numberwithin{equation}{section} 
\numberwithin{theorem}{section}
\numberwithin{lemma}{section}
\numberwithin{proposition}{section}
\numberwithin{corollary}{section}
\numberwithin{remark}{section}
\numberwithin{assumption}{section}

\DeclarePairedDelimiter\abs{\lvert}{\rvert}%
\DeclarePairedDelimiter\norm{\lVert}{\rVert}%
\makeatletter
\let\oldabs\abs
\def\abs{\@ifstar{\oldabs}{\oldabs*}}
\let\oldnorm\norm
\def\norm{\@ifstar{\oldnorm}{\oldnorm*}}
\makeatother

\newcommand{\E}{\mathbb{E}}
\newcommand{\Prob}{\mathbb{P}}
\newcommand{\indep}{\perp \!\!\! \perp}
\DeclareMathOperator{\Bprob}{pr}
\DeclareMathOperator{\diag}{diag}
\DeclareMathOperator{\outlier}{out}
\DeclareMathOperator{\PF}{pf}
\DeclareMathOperator{\DP}{DP}
\DeclareMathOperator{\load}{(L)}
\DeclareMathOperator{\factor}{(G)}

\DeclareMathOperator{\direct}{(\pi)}
\DeclareMathOperator{\directvar}{(\varphi^2)}
\DeclareMathOperator{\V}{Var}
\DeclareMathOperator{\lfsr}{lfsr}
\DeclareMathOperator{\mvec}{vec}
\DeclareMathOperator{\isim}{\stackrel{i.i.d}{\sim}}

\begin{document}

\title{A statistical framework for GWAS of high dimensional phenotypes using summary statistics, with application to metabolite GWAS}

\author{Weiqiong Huang$^{1}$, Emily C. Hector$^{2}$, Joshua Cape$^{3}$, Chris McKennan$^{1,}\thanks{To whom correspondence should be addressed (chm195@pitt.edu).}$\\[4pt]
\textit{$^1$Department of Statistics,
University of Pittsburgh}
\\
\textit{$^2$Department of Statistics, North Carolina State University}\\
\textit{$^3$Department of Statistics, University of Wisconsin}\\[2pt]
}
\maketitle

\begin{abstract}
The recent explosion of genetic and high dimensional biobank and `omic' data has provided researchers with the opportunity to investigate the shared genetic origin (pleiotropy) of hundreds to thousands of related phenotypes. However, existing methods for multi-phenotype genome-wide association studies (GWAS) do not model pleiotropy, are only applicable to a small number of phenotypes, or provide no way to perform inference. To add further complication, raw genetic and phenotype data are rarely observed, meaning analyses must be performed on GWAS summary statistics whose statistical properties in high dimensions are poorly understood. We therefore developed a novel model, theoretical framework, and set of methods to perform Bayesian inference in GWAS of high dimensional phenotypes using summary statistics that explicitly model pleiotropy, beget fast computation, and facilitate the use of biologically informed priors. We demonstrate the utility of our procedure by applying it to metabolite GWAS, where we develop new nonparametric priors for genetic effects on metabolite levels that use known metabolic pathway information and foster interpretable inference at the pathway level. 
\end{abstract}


\section{Introduction}
\label{section:Introduction}
Genome-wide association studies (GWAS) investigate the relationship between the genotype at single nucleotide polymorphisms (SNPs) and phenotypes, and have become an essential tool for deciphering the genetic basis of human variation \citep{GWAS_overview}. Recently, the explosion of high dimensional biobank and `omic' data have presented biologists with the opportunity to understand the common genetic origin (pleiotropy) of hundreds to thousands of related phenotypes \citep{MultiUK2,Example_Brain,Metsim,Example_Microbiome}. Of particular interest are GWAS of metabolites, small molecules that are by- or end-products of cell metabolism. Metabolites are crucial to establishing functional links between genotype and disease \citep{MetabDisease}; thus, metabolite GWAS (mtGWAS) have the potential to transform our understanding of the etiology of disease and uncover new biomarkers for targeted therapies \citep{mtGWAS_pleiotropy}. 

The many unsolved statistical and methodological challenges in mtGWAS highlight the difficulties of performing GWAS with high dimensional phenotypes. First, metabolite levels exhibit systemic genetic and non-genetic correlation \citep{MetabMiss,Metsim}. However, existing methods for multi-phenotype GWAS that incorporate inter-phenotype correlations can only accommodate a small number of phenotypes \citep{Qi-etal,Dai-etal,Liu-etal,Lu-etal}, can only test whether a SNP is related to at least one phenotype \citep{Brain_Multi,MultiUK}, or provide no inferential guarantees \citep{MultiUK2,MultiTrait_LowRank}. Consequently, nearly all mtGWAS studies ignore pleiotropy and perform pairwise marginal regressions with a stringent Bonferroni correction to adjust for the \#SNPs $\times$ \#metabolites $\gtrsim 10^8$ tests. Second, raw metabolite and genetic data are rarely available due to privacy concerns. Instead, practitioners typically only have access to summary statistics from the \#SNPs $\times$ \#metabolites marginal regressions of metabolite levels onto SNP genotypes. Although existing approaches have treated SNPs as analogous to biological samples and applied standard factor analysis tools to analyze the matrix of summary statistics \citep{MultiUK2,MultiTrait_LowRank}, we show these approaches lead to erroneous estimates and inference and that new methods and theoretical results are required. Lastly, substantial biological information on the grouping of metabolites into biological pathways should be leveraged to improve power and interpretability. The latter is essential, since deciphering results from mtGWAS with thousands of metabolites is exceedingly challenging, whereas interpreting functional pathways is far simpler.

\begin{figure}
\centering
\includegraphics[width=0.5\textwidth]{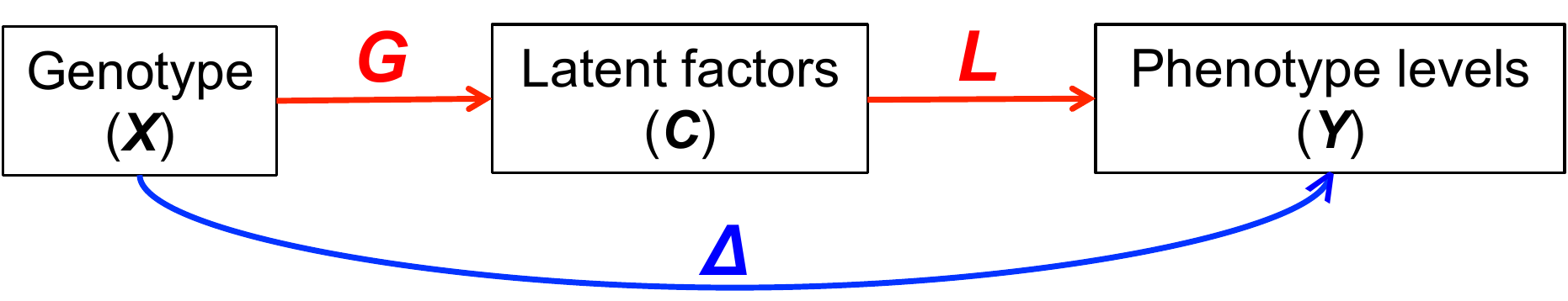}
\caption{A graphical model for the effect of genotype on phenotype levels. Red: indirect effect; blue: direct effect. Parameters above arrows are the parameters of interest.}
\label{figure:GraphModel}
\end{figure}

To address these critical obstacles, we develop a new statistical model, theoretical framework, and methods for GWAS with high dimensional phenotypes. Figure~\ref{figure:GraphModel} gives a graphical depiction of our model, which decomposes the total genetic effect into an indirect effect mediated by a set of latent factors and a direct effect. Factors represent potentially genetically determined biological processes, such as glucose metabolism in metabolomics or gut architecture in microbiomics \citep{MicrobiomeExample2}, that regulate the levels of many phenotypes and consequently help model pleiotropy and non-genetic correlation. The direct effect captures remaining idiosyncratic genetic variation. 

Our theoretical framework is essential because our estimates rely on performing factor analysis with GWAS summary statistics whose statistical behavior is, to our knowledge, currently unknown. We therefore develop, to our knowledge, the first set of theoretical guarantees for factor analysis with summary statistics. These are more than mere technical developments, as we show that ostensibly trivial dependencies between SNP genotypes cause estimators' asymptotic distributions to differ substantially from those derived from raw data.

We lastly develop two new methods to estimate the indirect and direct genetic effects in Figure~\ref{figure:GraphModel} that attempt to solve the abovementioned methodological challenges. The first is a method to estimate the number of latent factors we call dBEMA (\ul{d}ependent \ul{B}ulk \ul{E}igenvalue \ul{Ma}tching), which accounts for unobserved dependencies between SNP genotypes that cause existing methods designed for raw data to drastically misestimate the number of factors when applied to GWAS summary statistics. The second is a general Bayesian factor analysis method called HiGSS (\underline{H}igh D\underline{i}mensional \underline{G}WAS with \underline{S}ummary \underline{S}tatistics), which performs theoretically justified Bayesian inference on model parameters. Unlike traditional computationally onerous posterior sampling procedures and sub-optimal variational approximations, HiGSS leverages our abovementioned theoretical results to efficiently determine posteriors via empirical Bayes without sacrificing statistical fidelity. Critically, HiGSS facilitates the use of biologically-informed priors that would be computationally intractable in traditional Bayesian factor analysis pipelines, such as phylogenetic and spatial priors in microbiome and brain image data \citep{MicrobiomePrior,BrainPrior}. While we design HiGSS to be compatible with general priors, we devote considerable efforts to constructing appropriate nonparametric priors for the direct and indirect effects in mtGWAS, which leverage known metabolic pathway information to facilitate powerful and interpretable inference at the pathway level.

The remainder of this paper is organized as follows. We describe our model for summary statistics in Section \ref{section:Model}. The proposed estimation algorithms and their theoretical guarantees are presented in Section \ref{section:Method}. In Section \ref{section:mtGWAS}, we construct nonparametric priors for indirect and direct effects in mtGWAS. Section \ref{section:Simulations} investigates the finite sample performance of our proposed estimation approach. We perform a mtGWAS in Section \ref{section:RealData} to illustrate the application of our method. Section \ref{section:Conclusions} concludes.


\section{A model for the data}
\label{section:Model}

\subsection{Problem Notation}
\label{subsection:Notation}
For $\bm{Z} \in \mathbb{R}^{n \times m}$, we let $\bm{Z}_{i*}\in \mathbb{R}^m$, $\bm{Z}_{*j}\in \mathbb{R}^n$, and $\bm{Z}_{ij} \in \mathbb{R}$ be the $i$th row, $j$th column, and $(i,j)$th element of $\bm{Z}$. We use standard matrix normal notation, i.e. for $\bm{M}\in\mathbb{R}^{n \times m}$, $\bm{U} \in \mathbb{R}^{n \times n}$, and $\bm{V} \in \mathbb{R}^{m \times m}$, $\bm{Z} \sim MN(\bm{M},\bm{U},\bm{V})$ if $\mvec(\bm{Z}) \sim N( \mvec(\bm{M}),\bm{V}\otimes \bm{U} )$. Vector $\bm{x} \sim (\bm{\mu},\bm{A})$ if $\E(\bm{x})=\bm{\mu}$ and $\V(\bm{x})=\bm{A}$.

\subsection{A model for the unobserved genotype and phenotype data}
\label{subsection:UnobservedModel}
Let $\bm{Y} \in \mathbb{R}^{N \times M}$ be the unobserved levels of $M$ phenotypes (metabolites) measured in $N$ unrelated individuals and $\bm{X} \in \mathbb{R}^{N \times S}$ be the unobserved genotypes at $S$ SNPs in the same individuals. Assume
\begin{assumption}
\label{assumption:X}
$\bm{X}$ is a random matrix with bounded, mean zero, and independent entries whose rows are identically distributed and $\V(\bm{X}_{is}) \geq \epsilon$ for some constant $\epsilon>0$ and all $i\in \{1,\ldots,n\}=[n]$ and $s \in [S]$.
\end{assumption}
The mean 0 and independence assumption are equivalent to mean-centering $\bm{X}$'s columns and pruning SNPs for linkage disequilibrium, where pruning helps avoid inferring redundant genetic effects. Figure~\ref{figure:GraphModel} provides a graphical description of our model for the effect of genotypes on phenotypes, which decomposes the total effect into a direct effect and an indirect effect mediated by $K$ latent factors $\bm{C} \in \mathbb{R}^{N \times K}$, and is equivalent to the following model: 
\begin{align}
\label{equation:Model}
\begin{aligned}
\bm{Y}_{N \times M} =& \bm{C}_{N \times K}\bm{L}_{M \times K}^{\top} + \bm{X}_{N \times S}\bm{\Delta}_{S \times M} + \bm{E}_{N \times M}, \quad \bm{E} \sim MN(0, I_N, \diag(\sigma_1^2,\ldots,\sigma_M^2))\\
\bm{C}_{N \times K}=& \bm{X}_{N \times S}\bm{G}_{S \times K} + \bm{\Xi}_{N \times K}, \quad \bm{\Xi}_{1*},\ldots,\bm{\Xi}_{N*} \isim (0,\bm{\Psi}_{K \times K}),
\end{aligned}
\end{align}
where $\bm{L}_{mk}$ is the effect of the $k$th latent factor on the $m$th phenotype and $\bm{G}_{sk}$ is the effect of the $s$th SNP on the $k$th factor. Under this model, $\bm{G}$, $\bm{L}$, and $\bm{\Delta}$ are the parameters of interest and $\E(\bm{Y} \mid \bm{X},\bm{G},\bm{L},\bm{\Delta}) = \bm{X}(\bm{G}\bm{L}^{\top} + \bm{\Delta})$ is the total genetic effect, where $\bm{G}_{s*}^{\top} \bm{L}_{m*}$ and $\bm{\Delta}_{sm}$ are the indirect and direct effects of the $s$th SNP on the $m$th phenotype. The non-genetic covariance between phenotypes $\V(\bm{Y}_{i*} \mid \bm{X},\bm{G},\bm{L},\bm{\Delta}) = \bm{L}\bm{\Psi}\bm{L}^{\top} + \diag(\sigma_1^2,\ldots,\sigma_M^2)$ is low rank plus diagonal, which is a standard assumption in metabolite, microbiome, and other multi-phenotype data \citep{MicroMetab,MetabMiss,MultiUK2}. Since we propose a Bayesian method to infer $\bm{G}$, $\bm{L}$, and $\bm{\Delta}$, we explicitly condition on them in the above and below probability statements to avoid confusion.

Two features of \eqref{equation:Model} help motivate assumptions on $\bm{G}$ and $\bm{\Delta}$. First, the first term in the total genetic coefficient $\bm{G}\bm{L}^{\top}+\bm{\Delta}$ captures patterns of inter-phenotype and inter-SNP effect sharing, suggesting $\bm{\Delta}$ should reflect idiosyncratic genetic variation. Second, both $\bm{G}$ and $\bm{\Delta}$ represent direct effects of genotype on a set of phenotypes, and are consequently expected to be sparse with small non-zero entries \citep{GWASEffectSizes}. We therefore assume throughout that
\begin{assumption}
\label{assumption:GDelta}
For some positive integer $A$ and constants $c,r\geq 0$
\begin{enumerate}[label=(\alph*)]
\item $\bm{\Delta}$ is a random matrix with independent entries that satisfy $\Prob(\bm{\Delta}_{sm} \neq 0)=\pi_{sm}$ and $\bm{\Delta}_{sm} \mid (\bm{\Delta}_{sm} \neq 0) \sim \sum_{a=1}^{A} \gamma_{sma}N(0,\tau_{sma}^2)$, where $\pi_{sm},\gamma_{sma} \in [0,1]$ and $\sum_{a=1}^{A}\gamma_{sma}=1$. Further, $\max_{s\in[S];m \in [M]; a\in [A]}(N^{1/2} \tau_{sma}) \leq c$ and $\max_{s\in[S];m \in [M]}\pi_{sm}=o(1)$ as $S,M \to \infty$.\label{Sparse:Delta}
\item $\max_{s\in[S]}\norm*{N^{1/2} \bm{G}_{s*}}_2 = O_P( \log^r N )$ and $S^{-1}\sum_{s=1}^S \norm*{N^{1/2} \bm{G}_{s*}}_2^2 = o_P(1)$ as $S,N \to \infty$.\label{Sparse:G}
\end{enumerate}
\end{assumption}
$\bm{\Delta}$'s entry-wise independence implies it captures idiosyncratic genetic variation which, along with its entries' symmetric spike and slab distributions, are classic assumptions on genetic effects in GWAS with one and more than one phenotype \citep{BSLMM,zhu2017bayesian,Biostat_metab,GWASEffectSizes,MTAG,Multi_SS}. The mixture-normal slab is quite general and can approximate a large class of symmetric distributions \citep{stephens2017false}. The parameters $\tau_{sma}$ determine the magnitude of $\bm{\Delta}$'s non-zero effects, where the conditions on $N^{1/2}\tau_{sma}$ and $N^{1/2}\bm{G}_{s*}$ reflect the observation that GWAS z-scores, whose magnitudes scale with $N^{1/2}$ times the SNPs' effect sizes, are typically small to moderate in real data \citep{zhu2017bayesian,GWASEffectSizes}. Sparsity is implied by our condition on $\pi_{sm}$ in~\ref{Sparse:Delta} and the sum of squares constraint in~\ref{Sparse:G}. It is important to note that while we assume $\bm{\Delta}$ is random, weak convergence results for all frequentist estimators in Section~\ref{subsection:theory} are stated conditional on $\bm{\Delta}$. Assuming $\bm{\Delta}$ is random can be thought of as a way to only consider $\bm{\Delta}$'s likely to arise in genetic data and ignore unrealistic ``corner cases''. We discuss assumptions for \eqref{equation:Model}'s remaining parameters in Section~\ref{subsection:theory}.


\subsection{Observed GWAS summary statistics}
\label{subsection:SSModel}
We assume the GWAS summary statistics $\hat{\beta}_{sm} = (\bm{X}_{*s}^{\top}\bm{X}_{*s})^{-1}\bm{X}_{*s}^{\top} \bm{Y}_{*m}$ from the regressions of the $m$th phenotype onto $s$th SNP's genotype are observed by the user. We also assume $\bm{X}_{*s}^{\top}\bm{X}_{*s}$ is known, although this can be estimated from public data if unavailable \citep{ThousandGenomes}. We define the observable matrix of standardized GWAS statistics $\hat{\bm{B}} \in \mathbb{R}^{S \times M}$ to be $\hat{\bm{B}}_{sm}= (\bm{X}_{*s}^{\top}\bm{X}_{*s})^{1/2} \hat{\beta}_{sm}$ which, for $\bm{D}=\diag\{ (\bm{X}_{*1}^{\top}\bm{X}_{*1}),\ldots,(\bm{X}_{*S}^{\top}\bm{X}_{*S}) \}$ and $\bm{R} = \bm{D}^{-1/2}\bm{X}^{\top}\bm{X} \bm{D}^{-1/2}$ (the in-sample correlation between SNPs), can be expressed as
\begin{align}
\label{equation:Bhat}
\begin{aligned}
&\hat{\bm{B}} = \bm{D}^{-1/2}\bm{X}^{\top}\bm{Y} = \tilde{\bm{G}}\bm{L}^{\top} + \bm{R}(\bm{D}^{1/2}\bm{\Delta}) + \tilde{\bm{E}}\\
&\tilde{\bm{G}} = \bm{R}(\bm{D}^{1/2}\bm{G}) + \bm{D}^{-1/2}\bm{X}^{\top}\bm{\Xi}, \quad \tilde{\bm{E}} \mid \bm{X} \sim MN(0, \bm{R}, \diag(\sigma_1^2,\ldots,\sigma_M^2)).
\end{aligned}
\end{align}
Standardizing summary statistics by $(\bm{X}_{*s}^{\top}\bm{X}_{*s})^{1/2}$ is akin to working with z-scores and begets rows (SNPs) with identical covariances, where $\bm{D}^{1/2}\bm{G}$ and $\bm{D}^{1/2}\bm{\Delta}$ are the standardized analogues of $\bm{G}$ and $\bm{\Delta}$. Although the diagonal elements of $\bm{D}$ grow with the sample size, Assumption~\ref{assumption:GDelta} implies $\bm{D}^{1/2}\bm{G}$ and $\bm{D}^{1/2}\bm{\Delta}$ are moderate. Evidently, $\hat{\bm{B}}$ depends on $\bm{D}^{1/2}\bm{G}$ and $\bm{D}^{1/2}\bm{\Delta}$ through $\bm{R}(\bm{D}^{1/2}\bm{G})$ and $\bm{R}(\bm{D}^{1/2}\bm{\Delta})$. However, since the off-diagonal elements of $\bm{R}$ are small under Assumption~\ref{assumption:X}, the sparsity assumptions in Assumption~\ref{assumption:GDelta} imply $\bm{R}(\bm{D}^{1/2}\bm{G})\approx \bm{D}^{1/2}\bm{G}$ and $\bm{R}(\bm{D}^{1/2}\bm{\Delta})\approx \bm{D}^{1/2}\bm{\Delta}$.

\subsection{Parameter identifiability}
The parameters of interest $(\bm{G},\bm{L})$ in \eqref{equation:Bhat} are not identifiable, since $(\bm{G}\bm{A},\bm{L}\bm{A}^{-\top})$ will give an identical likelihood for any invertible $K \times K$ matrix $\bm{A}$. We address this with the following proposition.
\begin{proposition}
\label{proposition:Gtilde}
If Assumptions~\ref{assumption:X}, \ref{assumption:GDelta}, and \ref{assumption:L} in Section~\ref{subsection:theory} hold, then for all large $N,M,S$:
\begin{enumerate}[label=(\roman*)]
    \item There exists a $(\bm{G},\bm{L})$ so that $\E(S^{-1}\tilde{\bm{G}}^{\top}\tilde{\bm{G}} \mid \bm{G})=I_K$, $M^{-1}\bm{L}^{\top}\bm{L}=\diag(\lambda_1,\ldots,\lambda_K)$ and $\lambda_1> \cdots > \lambda_K >0$. Further, the columns of $\bm{G}$ and $\bm{L}$ are unique up to sign.
    \item If $c^{-1} I_K \preceq E(S^{-1}\tilde{\bm{G}}^{\top}\tilde{\bm{G}} \mid \bm{G}) \preceq c I_K$ for some constant $c>1$, then for all $\epsilon \in (0,1)$, there exists a $\delta=\delta({\epsilon})>0$ such that $\Prob(\abs*{\tilde{\bm{G}}_{sk}}\geq \delta \mid \bm{G})\geq 1-\epsilon$ for all $s \in [S]$, $k \in [K]$. 
\end{enumerate}
\end{proposition}

We use the parametrization in (i) for the remainder of the manuscript, which is a classic way to identify components of factor models \citep{BCconf}. The orthogonality of the columns of $\tilde{\bm{G}}$ and $\bm{L}$ also help motivate Algorithm~\ref{algorithm:HIGSS} below, which uses singular value decomposition to derive estimates for $\bm{G}$, $\bm{L}$, and $\bm{\Delta}$. Part (ii) of Proposition~\ref{proposition:Gtilde} suggests that regardless of parametrization, $\tilde{\bm{G}}$ will never be sparse even if $\bm{G}$ is sparse. This follows because $\tilde{\bm{G}}$ in \eqref{equation:Bhat} is a noisy version of $\bm{G}$, whose error $\bm{D}^{-1/2}\bm{X}^{\top}\bm{\Xi}$ is dense. This is in stark contrast with existing methods that assume a sparse low rank structure on $\hat{\bm{B}}$ \citep{MultiTrait_LowRank}, suggesting they enforce non-existent sparsity.

\section{A general method to estimate model parameters}
\label{section:Method}

\subsection{HiGSS: an algorithm to estimate indirect and direct effects}
\label{subsection:Algorithm}
Here we present a Bayesian method we call HiGSS (\underline{H}igh D\underline{i}mensional \underline{G}WAS with \underline{S}ummary \underline{S}tatistics) in Algorithm~\ref{algorithm:HIGSS} below to derive the posterior $\Bprob(\bm{G},\bm{L},\bm{\Delta} \mid \hat{\bm{B}},\bm{D})$ assuming the number of latent factors $K$ is known. To circumvent the computational and statistical issues that arise in Bayesian factor analysis, and as alluded to by Proposition~\ref{proposition:Gtilde}, HiGSS first uses singular value decomposition to derive frequentist estimates for the parameters of interest and subsequently leverages the asymptotic theory developed Section~\ref{subsection:theory} to calculate the posterior. HiGSS assumes the priors on $\bm{G}$, $\bm{L}$, and $\bm{\Delta}$ are independent. 




\begin{Algorithm}[HiGSS]
\label{algorithm:HIGSS}
\normalfont
\textit{Input data}: Standardized GWAS estimates $\hat{\bm{B}} \in \mathbb{R}^{S \times M}$, $\bm{D}=\diag\{ (\bm{X}_{*1}^{\top}\bm{X}_{*1}),\allowbreak \ldots, \allowbreak$ $ (\bm{X}_{*S}^{\top}\bm{X}_{*S}) \}$, number of latent factors $K \geq 0$, and sample size $N$.\\
\textit{Input priors}: $\Bprob(\bm{G} \mid \bm{D})=\prod_{s=1}^S \Bprob_s(\bm{G}_{s*} \mid \bm{D})$, $\Bprob(\bm{L} \mid \bm{D})=\prod_{m=1}^M \Bprob_m(\bm{L}_{m*})$, and $\bm{\Delta}$-specific hyperparameters $ \{\{\pi_{sm},\{ (\gamma_{sma},\tau_{sma}^2) \}_{a\in[A]}\}\}_{s\in[S];m\in[M]}$.\\
\textit{Output}: Posterior distribution $\Bprob(\bm{G},\bm{L},\bm{\Delta} \mid \hat{\bm{B}},\bm{D})$

\begin{enumerate}
\item Let $\bm{\Gamma}=\diag(\gamma_1,\ldots,\gamma_{K})$, $\bm{U} \in \mathbb{R}^{S \times K}$, and $\bm{V} \in \mathbb{R}^{M \times K}$ be $\hat{\bm{B}}$'s first $K$ singular values and left and right singular vectors. Define $\hat{\bm{G}} = S^{1/2}\bm{U}$, $\hat{\bm{L}} = S^{-1/2}\bm{V}\bm{\Gamma}$, and $\hat{\bm{\Delta}} = \hat{\bm{B}} - \hat{\bm{G}}\hat{\bm{L}}^{\top}$ to be estimators for $\bm{G}$, $\bm{L}$, and $\bm{\Delta}$.\label{HiGSS:Est}
\item Calculate $\Bprob(\bm{G} \mid \hat{\bm{G}},\bm{D}) \propto \prod_{s=1}^S \mathcal{N}( \hat{\bm{G}}_{s*}; \bm{D}_{ss}^{1/2}\bm{G}_{s*},I_K ) \Bprob_s(\bm{G}_{s*} \mid \bm{D})$, where $\mathcal{N}(\bm{x};\bm{\mu},\bm{V})$ is the density at $\bm{x}$ of a normal with mean $\bm{\mu}$ and variance $\bm{V}$.\label{HiGSS:Ghat} 
\item Let $\hat{\sigma}_m^2 = S^{-1}\sum_{s=1}^S \hat{\bm{\Delta}}_{sm}^2$ be an estimate for $\sigma_m^2$. Calculate $\Bprob(\bm{L} \mid \hat{\bm{L}},\bm{D}) \propto \prod_{m=1}^M \mathcal{N}\{\hat{\bm{L}}_{m*}; \bm{L}_{m*},\allowbreak (N^{-1}+S^{-1})\hat{\sigma}_m^2 I_K \} \Bprob_m(\bm{L}_{m*})$.\label{HiGSS:Lhat} 
\item Let $\Bprob(\bm{\Delta} \mid \hat{\bm{\Delta}},\bm{D}) \propto \prod_{m=1}^M \prod_{s=1}^S \mathcal{N}(\hat{\bm{\Delta}}_{sm}; \bm{D}_{ss}^{1/2}\bm{\Delta}_{sm},\hat{\sigma}_m^2)\Bprob(\bm{\Delta}_{sm} \mid \pi_{sm},\tau_{sm}^2)$, where $\bm{\Delta}_{sm} \mid ( \pi_{sm},\tau_{sm}^2 ) \sim (1-\pi_{sm}) \delta_0 + \pi_{sm}\sum_{a=1}^A \gamma_{sma} N(0,\tau_{sma}^2)$.\label{HiGSS:Delta} 
\item Return the posterior $\Bprob(\bm{G},\bm{L},\bm{\Delta} \mid \hat{\bm{B}},\bm{D}) = \Bprob(\bm{G} \mid \hat{\bm{G}},\bm{D}) \Bprob(\bm{L} \mid \hat{\bm{L}},\bm{D}) \Bprob(\bm{\Delta} \mid \hat{\bm{\Delta}},\bm{D})$.\label{HiGSS:Post}
\end{enumerate}
\end{Algorithm}

The intuition for step~\ref{HiGSS:Est} is if $\bm{\Delta}$ is sparse, \eqref{equation:Bhat} and the accuracy of singular value decomposition suggest $\hat{\bm{G}} \approx \tilde{\bm{G}}$, $\hat{\bm{L}} \approx \bm{L}$, and the residuals $\hat{\bm{\Delta}}$ should be approximately $\bm{D}^{1/2}\bm{\Delta}$ plus independent noise \citep{BCconf}. The subscripts ``$s$'' and ``$m$'' in $\bm{G}$'s and $\bm{L}$'s prior distributions indicate these priors may depend on SNP $s$- and phenotype $m$-specific parameters, and we justify the likelihoods
for $\hat{\bm{G}}$, $\hat{\bm{L}}$, and $\hat{\bm{\Delta}}$ from steps~\ref{HiGSS:Ghat}-\ref{HiGSS:Delta} in Section~\ref{subsection:theory}. Since $\bm{\Delta}$ contains $S \times M\gtrsim 10^8$ elements, computing its posterior is intractable unless an appropriate prior is chosen. The spike and mixture-normal slab prior in step~\ref{HiGSS:Delta}, whose parameters' definitions match those in Assumption~\ref{assumption:GDelta}, is conjugate and begets tractable computation. Its concordance with Assumption~\ref{assumption:GDelta} is coincidental and is chosen to match the prevailing choices of priors in GWAS \citep{GWASEffectSizes}.

Step~\ref{HiGSS:Post} assumes the total posterior $\Bprob(\bm{G},\bm{L},\bm{\Delta} \mid \hat{\bm{B}},\bm{D})$ can be factored into the product of posteriors defined in the previous three steps. To see why this is appropriate, note $\{\hat{\bm{G}},\hat{\bm{L}},\hat{\bm{\Delta}}\}$ is a function of $\hat{\bm{B}}$ and $\hat{\bm{B}}=\hat{\bm{G}} \hat{\bm{L}}^{\top} + \hat{\bm{\Delta}}$, meaning the total posterior is equivalent to
\begin{align*}
    \Bprob(\bm{G},\bm{L},\bm{\Delta} \mid \hat{\bm{G}},\hat{\bm{L}},\hat{\bm{\Delta}},\bm{D}) \propto &\Bprob(\hat{\bm{G}},\hat{\bm{L}},\hat{\bm{\Delta}} \mid \bm{G},\bm{L},\bm{\Delta},\bm{D}) \Bprob(\bm{G}\mid\bm{D}) \Bprob(\bm{L}\mid\bm{D}) \Bprob(\bm{\Delta}\mid \bm{D})
\end{align*}
Step~\ref{HiGSS:Post} therefore follows provided $\hat{\bm{G}}$, $\hat{\bm{L}}$, and $\hat{\bm{\Delta}}$ are independent conditional on $\{\bm{G},\bm{L},\bm{\Delta},\bm{D}\}$, which we prove is true asymptotically in Section~\ref{subsection:theory}.

Perhaps the most unexpected component of Algorithm~\ref{algorithm:HIGSS} is $\hat{\bm{L}}$'s likelihood in step~\ref{HiGSS:Lhat}. Since $\hat{\bm{B}}$ has $S$ rows, existing singular value decomposition results suggest $\hat{\bm{L}}_{m*}$'s asymptotic variance should be $S^{-1} \hat{\sigma}_m^2 I_K$ because $\hat{\bm{L}}_{m*}$ is ostensibly estimated on $S$ degrees of freedom \citep{BCconf}. This is quite different from the variance in step~\ref{HiGSS:Lhat} which, because $N$ is typically smaller than $S$, is closer to $N^{-1} \hat{\sigma}_m^2 I_K$ than $S^{-1} \hat{\sigma}_m^2 I_K$. This is sensible because $\hat{\bm{B}}$ in \eqref{equation:Bhat} is obtained by projecting the $N$ samples up to the space of $S$ SNPs, implying $\hat{\bm{L}}_{m*}$ is actually estimated on something closer to $N$ degrees of freedom. Theorem~\ref{theorem:Lhat} and Corollary~\ref{corollary:Lhat} in Section~\ref{subsection:theory} below, which prove the veracity of $\hat{\bm{L}}$'s likelihood, are to our knowledge the first results of their kind and, as we show in Section~\ref{section:Simulations}, are critical to the fidelity of inference on $\bm{L}$.

Algorithm~\ref{algorithm:HIGSS} is general in that the likelihoods in steps~\ref{HiGSS:Ghat}-\ref{HiGSS:Delta} and form for the posterior in step~\ref{HiGSS:Post} are the same regardless of the phenotypes being analyzed. However, the choice of priors will likely depend on the type of phenotypes. For example, the set of priors chosen in metabolomic applications may be different that those in microbiome data. We therefore provide general recommendations for how to choose priors for $\bm{G}$ and $\bm{L}$ and $\bm{\Delta}$'s hyperparameters in Section~\ref{subsection:Priors}.

\subsection{Choice of priors}
\label{subsection:Priors}
First, recall $\bm{G}$ is the effect of genotype on a small number ($K$) of latent phenotypes. Since $\hat{\bm{G}}$'s likelihood is normal in Algorithm~\ref{algorithm:HIGSS}, we can either treat $\bm{G}$'s $K$ columns as independent and use the many single-phenotype GWAS priors designed for and normal data (e.g. \citet{zhu2017bayesian} and \citet{GWASEffectSizes}), or model $\bm{G}$'s columns dependencies using recent methods designed multi-phenotype GWAS \citep{MTAG,Multi_SS}. We take the former approach in our metabolomic application.

The likelihoods for $\hat{\bm{L}}$ and $\hat{\bm{\Delta}}$ in Algorithm~\ref{algorithm:HIGSS} suggest the simplest approach would be to treat estimating $\bm{L}$ and $\bm{\Delta}$ as a normal means problem and use empirical Bayes to derive column-specific and matrix-wide priors, respectively, which can be done efficiently using \citet{stephens2017false}. An alternative would be to incorporate prior knowledge on the relationships between the $M$ phenotypes into the prior on $\bm{L}$ and $\bm{\Delta}$'s hyperparameters. For example, it is straightforward to incorporate phylogenetics in $\bm{L}$'s prior using the methods discussed in \citet{MicrobiomePrior}, and we show in Section~\ref{section:mtGWAS} how to leverage metabolic pathways in our prior on $\bm{L}$ and $\bm{\Delta}$'s hyperparameters.

The likelihoods for $\hat{\bm{G}}$ and $\hat{\bm{L}}$ in Algorithm~\ref{algorithm:HIGSS} can be written as a product of likelihoods across their $K$ columns. Consequently, provided their priors also factor across their columns, posterior computation on $\bm{G}$ and $\bm{L}$ can be done in parallel across their columns, which is not true for traditional Bayesian or recent empirical Bayes-based factor analysis methods \citep{EBPCA,FLASH}. 

\subsection{Determining the number of latent factors}
\label{subsection:K}
Estimating $K$ is critical, since underestimating it can cause one to miss important genetic variation and overestimating it by too much can invalidate statistical guarantees. Given $\hat{\bm{B}}$'s model in \eqref{equation:Bhat}, it is tempting to reason that because the off-diagonals of $\bm{R}$ (the in-sample correlation between SNPs) are approximately zero, one can treat the entries of the error matrix $\tilde{\bm{E}}$ as independent and use one of the many estimators designed for omic data with independent errors \citep{owen2016bi,DPA,PA_theory,ke2021estimation}. Unfortunately, as shown in Section~\ref{section:Simulations}, ignoring the dependencies produced by $\bm{R}$'s off-diagonal elements causes these methods to fail.


One way to address the abovementioned issues could be to use \citet{CorrConf}, which explicitly models the dependence. However, applying their method in this setting would require knowing $\bm{R}$, which is unobservable. A second option might be to use methods whose theoretical results allow $\tilde{\bm{E}}$'s entries be dependent, such as \citet{bai2002determining}, \citet{EigRatio}, or \citet{onatski2010determining}. However, the former two require factor loadings be unrealistically large (i.e. satisfy the pervasive factor assumption) and the latter patently fails in simulated and real omic data \citep{FALCO}. To address these deficiencies, note that if we assume for simplicity that $\bm{\Delta}$ in \eqref{equation:Bhat} is 0,
\begin{align*}
    \hat{\bm{B}} = \underbrace{\tilde{\bm{G}}\bm{L}^{\top}}_{\text{rank-$K$ signal}} + \underbrace{\tilde{\bm{E}}}_{\text{noise}}.
\end{align*}
Consequently, any of $\hat{\bm{B}}$'s singular values that exceed those in a pure noise model must have arisen from the low rank signal. This suggests that if we knew $\tilde{\bm{E}}$'s maximum singular value $s_{\max}$, a natural estimator for $K$ is the number of $\hat{\bm{B}}$'s singular values that exceed $s_{\max}$. We use this reasoning to motivate our novel method dBEMA (\ul{d}ependent \ul{B}ulk \ul{E}igenvalue \ul{Ma}tching), which models $\hat{\bm{B}}$'s bulk spectrum to predict $\tilde{\bm{E}}$'s maximum singular value. We describe dBEMA below.

Our method is an extension of \citet{ke2021estimation} to the setting where $\tilde{\bm{E}}$ has dependent rows. To be consistent with Section~\ref{section:Model}, we assume throughout that $\bm{E}$ in \eqref{equation:Model}, and therefore $\tilde{\bm{E}}$ in \eqref{equation:Bhat}, is Gaussian. However, the universality theory of eigenvalues implied by \citet{onatski2010determining} indicates we can relax this to only require $\bm{E}$'s entries satisfy weaker moment conditions. 

To motivate the method, suppose $S=S(N)$ and $M=M(N)$ such that $S/N \to \gamma_S \in (0,\infty)$ and $M/N \to \gamma_M \in (0,\infty)$, and let $F_{R,N}$ and $F_{\sigma^2,N}$ be the empirical distributions of $\bm{R}$'s $S$ eigenvalues and $\{\sigma_m^2\}_{m \in [M]}$, respectively. If the distributions $\lim_{N \to \infty} F_{R,N} = F_R$ and $\lim_{N \to \infty} F_{\sigma^2,N} = F_{\sigma^2}$ exist and are supported on compact intervals, Lemmas 1 and 3 in \citet{onatski2010determining} imply $F_R$ and $F_{\sigma^2}$ determine the behavior of $\tilde{\bm{E}}$'s bulk and largest singular values. The problem therefore boils down to estimating $F_R$ and $F_{\sigma^2}$, since this will allow us to predict $\tilde{\bm{E}}$'s maximum singular value.

First, Assumption~\ref{assumption:X} implies $F_R$ is exactly the Marchenko–Pastur distribution \citep{MPlaw}, which for $x \geq 0$, $\lambda_- = (1-\sqrt{\gamma_S})^2$, and $\lambda_+= (1+\sqrt{\gamma_S})^2$ takes the form
\begin{align*}
    F_R(x) = (1 - \gamma_S^{-1})1\{\gamma_S>1\} + \mathop{\smallint}\limits_{0}^{x} f_{R}(t) dt, \quad f_{R}(t) = (2\pi \gamma_S t)^{-1}\sqrt{(\lambda_+ - t)(t-\lambda_-)}1\{t\in[\lambda_-,\lambda_+]\}.
\end{align*}
Unlike $F_R$, $F_{\sigma^2}$ is unknown and must be estimated from the data. Following \citet{ke2021estimation}, we model $F_{\sigma^2}$ as a gamma distribution and use $\hat{\bm{B}}$'s bulk singular values to estimate its parameters. Briefly, let $\hat{s}_{\min(N,M,S)}\leq \cdots \leq \hat{s}_{1}$ be $\hat{\bm{B}}$'s non-zero singular values and define $\mathcal{B} = \{\hat{s}_{\lfloor(\alpha/2) \min(N,M,S) \rfloor},\ldots,\hat{s}_{\lfloor(1-\alpha/2) \min(N,M,S) \rfloor}\}$ for some $\alpha \in (0,1)$ to be $\hat{\bm{B}}$'s bulk singular values. Since $\mathcal{B}$ excludes $\hat{\bm{B}}$'s $K$ largest singular values for sufficiently large $N$, $\mathcal{B}$'s elements are approximately the inner $1-\alpha$ fraction of $\tilde{\bm{E}}$'s singular values. Consequently, these can be predicted given $F_R$ and the rate and scale parameters that parameterize $F_{\sigma^2}$. We therefore minimize the difference between $\mathcal{B}$'s elements and their predicted values as a function of the rate and scale parameters to estimate $F_{\sigma^2}$. Equipped with $F_R$ and an estimate for $F_{\sigma^2}$, we lastly determine $\hat{F}_{s_{max}}$, an estimate for the distribution of $\tilde{\bm{E}}$'s maximum singular value, and define our estimate for the number of factors to be the number $\hat{\bm{B}}$'s singular values that exceed the $1-\beta$ quantile of $\hat{F}_{s_{max}}$ for some user-specified $\beta \in (0,1)$. The algorithm is given in Section~\ref{appendix:section:dBEMA}.

As mentioned above, the theoretical work used to motivate dBEMA requires $F_{\sigma^2}$ have bounded support, i.e. $\min\{x\in\mathbb{R} \cup \{\infty\}: F_{\sigma^2}(x)=1\}< \infty$, which belies our decision to model $F_{\sigma^2}$ as a gamma distribution. We use simulated data in Section~\ref{section:Simulations} to show dBEMA performs well despite this.





\subsection{Theoretical justifications}
\label{subsection:theory}
Here we provide a theoretical justification for the likelihoods for $\hat{\bm{G}}$, $\hat{\bm{L}}$, and $\hat{\bm{\Delta}}$ used in Algorithm~\ref{algorithm:HIGSS}. Theory for the consistency of our estimate for $K$ proposed in Section~\ref{subsection:K} will appear in a subsequent version of this manuscript. We first present an assumption on the parameters in \eqref{equation:Model}.

\begin{assumption}
\label{assumption:L}
Let $c>1$ and $\epsilon \in (0,1)$ be constants and define $\lambda_{K+1}=0$.
\begin{enumerate}[label=(\alph*)]
\item $N/S,M/S \in [c^{-1},c]$, $K \leq c$, $\sigma_m^2\in[c^{-1},c]$ for all $m\in[M]$, and $\bm{\Xi}_{1k}$ is sub-exponential for all $k \in[K]$.\label{L:SampleSize}
\item $\E(S^{-1}\tilde{\bm{G}}^{\top} \tilde{\bm{G}} \mid \bm{G}) = I_K$ and $M^{-1}\bm{L}^{\top}\bm{L}=\diag(\lambda_1,\ldots,\lambda_K) \succ 0$ such that $\lambda_{k}/\lambda_{k+1} \geq 1+c^{-1}$ and $\lambda_k \in [c^{-1}N^{\epsilon - 1}, c]$ for all $k \in [K]$ and $\abs*{\bm{L}_{mk}} \leq c \lambda_{k}^{1/2}$ for all $m\in[M]$ and $k \in [K]$.\label{L:Lambda}
\end{enumerate}
\end{assumption}
\begin{remark}
\label{remark:AsummptionL}
The conditions in \ref{L:Lambda} implicitly assume $\bm{G}$ and $\bm{L}$ are non-random, although we continue to condition on them in all probability statements below to avoid any confusion. It is straightforward to replace \ref{L:Lambda} we an equivalent assumption that allows $\bm{G}$ and $\bm{L}$ to be random.
\end{remark}

We assume $N$, $M$, and $S$ are comparable in \ref{L:SampleSize} to be consisted with `omic' and biobank data. The requirement that $\E(S^{-1}\tilde{\bm{G}}^{\top} \tilde{\bm{G}} \mid \bm{G})=I_K$ and $M^{-1}\bm{L}^{\top}\bm{L}$ be diagonal with decreasing diagonal elements is without loss of generality by Proposition~\ref{proposition:Gtilde}, where our condition on the ratio of successive diagonal elements is a standard eigengap assumption and ensures we can identify the columns of $\bm{L}$ and $\bm{G}$ up to sign \citep{BCconf}. The parameter $\lambda_k$ is the average sum of squares of the $k$th loading $\bm{L}_{*k}$ and is interpretable as the $k$th latent factor's signal strength, where the ``pervasive factor assumption'' is satisfied if $\lambda_1,\ldots,\lambda_K \asymp 1$ \citep{Pervasive}. While common in the literature, the pervasive factor assumption is patently violated in `omic' data, where methods that utilize it often cannot recover non-pervasive, but critical, sources of omic variation \citep{FALCO}. This is especially important in our data application whose most genetically regulated factors are weaker. We therefore allow $\lambda_k$ to be large ($\lambda_k \asymp 1$, pervasive) or moderate to small ($\lambda_k<<1$, non-pervasive) to ensure we can perform inference on as many types of factors as possible.

We first justify $\hat{\bm{G}}'s$ likelihood in step~\ref{HiGSS:Ghat} of Algorithm~\ref{algorithm:HIGSS}. We define $\mathscr{C}_d$ and $\Phi_d$ to be the set of all convex sets and the distribution function for the standard normal in $\mathbb{R}^d$, respectively, for the remainder of the section.

\begin{theorem}
\label{theorem:Ghat}
Suppose Assumptions~\ref{assumption:X}, \ref{assumption:GDelta}, and \ref{assumption:L} hold. Then as $N,M,S\to \infty$, there exist $a_1,\ldots,a_K \in \{-1,1\}$ so that for any fixed set $\mathcal{S} \subset [S]$,
\begin{align*}
&\sup_{\substack{ R_s \in \mathscr{C}_K }}\abs{ \Prob\left[ \mathop{\bigcap}\limits_{s \in \mathcal{S}} \{\hat{\bm{G}}_{s *}-\bm{D}_{s s}^{1/2}\diag(a_1,\ldots,a_K)\bm{G}_{s *}\} \in R_s  \mid \bm{G},\bm{L},\bm{\Delta},\bm{X} \right] - \prod_{s \in \mathcal{S}}\Phi_K(R_s) } = o_P(1).
\end{align*}
\end{theorem}
Since the sign multipliers $a_1,\ldots,a_K$ are the same for each SNP $s$, this shows we can estimate each column of $\bm{G}$ up to sign. Theorem~\ref{theorem:Ghat} also justifies treating $\hat{\bm{G}}$'s rows in Algorithm~\ref{algorithm:HIGSS} as independent standard normals. We next justify step~\ref{HiGSS:Lhat} of Algorithm~\ref{algorithm:HIGSS}.

\begin{theorem}
\label{theorem:Lhat}
Suppose Assumptions~\ref{assumption:X}, \ref{assumption:GDelta}, and \ref{assumption:L} hold and $a_1,\ldots,a_K$ are as given in Theorem~\ref{theorem:Ghat}. Then the following hold as $N,M,S \to \infty$:
\begin{enumerate}[label=(\roman*)]
\item $\hat{\sigma}_m^2 = \sigma_m^2 + o_P(1)$ for all $m \in [M]$.\label{theorem:Lhat:sigma2}
\item $\mathop{\sup}\limits_{R \in \mathscr{C}_1}\abs{ \Prob\left\{ \norm*{ \hat{\bm{L}}_{m*} }_2^{-1}( \hat{\bm{G}}_{s*}^{\top}\hat{\bm{L}}_{m*} - \bm{D}_{ss}^{1/2}\bm{G}_{s*}^{\top}\bm{L}_{m*} ) \in R \mid \bm{G},\bm{L},\bm{\Delta},\bm{X} \right\} - \Phi_1(R) } = o_P(1)$ for any $s\in[S]$ and $m\in[M]$ provided $N^{1/2}\norm*{\bm{L}_{m*}}_2 \to \infty$.\label{theorem:Lhat:GL}
\item $\mathop{\sup}\limits_{\substack{ R_m \in \mathscr{C}_K }}\abs{ \Prob\left[ \mathop{\bigcap}\limits_{m \in \mathcal{M}} \{(N^{-1}+S^{-1})\hat{\sigma}_m^2\}^{-1/2}(\hat{\bm{L}}_{m *}-\bm{A}\bm{L}_{m *}) \in R_m \mid \bm{G},\bm{L},\bm{\Delta},\bm{X} \right] - \prod\limits_{m \in \mathcal{M}}\Phi_K(R_m) }\allowbreak = o_P(1)$ for any fixed set $\mathcal{M} \subset [M]$ and some $\bm{A} \in \mathbb{R}^{K \times K}$ such that $\norm*{ \bm{A} - \diag(a_1,\ldots,a_K) }_2=O_P(N^{-1/2})$.\label{theorem:Lhat:L}
\end{enumerate}
\end{theorem}
Results~\ref{theorem:Lhat:sigma2} and \ref{theorem:Lhat:GL} imply $\hat{\sigma}_m^2$ in step~\ref{HiGSS:Lhat} of Algorithm~\ref{algorithm:HIGSS} is consistent for $\sigma_m^2$ and we can infer indirect effects $\bm{G}_{s*}^{\top}\bm{L}_{m*}$. The condition $N^{1/2}\norm*{\bm{L}_{m*}}_2 \to \infty$ will hold if phenotype $m$ is non-trivially affected by at least one latent factor, and appears to be satisfied for nearly all $m$ in our data application. Result~\ref{theorem:Lhat:L} implies $\hat{\bm{L}}_{m*}$ is asymptotically normal with asymptotic variance as given in step~\ref{HiGSS:Lhat}. However, $\hat{\bm{L}}_{m*}$'s asymptotic distribution is centered around $\bm{A}\bm{L}_{m*}$ rather than $\diag(a_1,\ldots,a_K)\bm{L}_{m*}$, which is a standard property of singular value decomposition-based estimates \citep{CATE}. Although their difference is small, it may not be small enough to guarantee we can infer $\bm{L}_{m*}$'s sparsity patterns, since $(\bm{A}\bm{L}_{m*})_{k}$ may be larger than its estimate's standard deviation even if $\bm{L}_{mk}=0$. This is especially important when data contain both pervasive and non-pervasive factors, since $\bm{A}$'s off-diagonal elements risk contaminating sparser, non-pervasive factor loadings with dense pervasive factor loadings. Corollary~\ref{corollary:Lhat} below shows that no such contamination occurs.

\begin{corollary}
\label{corollary:Lhat}
In addition to the conditions of Theorem~\ref{theorem:Lhat}, suppose $\lambda_k \geq \eta$ or $\lambda_k \to 0$ as $M\to\infty$ for all $k \in [K]$ and some constant $\eta>0$. Define the number of pervasive factors $K_{\PF}$ to be 0 if $\lambda_1 \to 0$ and $K_{\PF}\in [K]$ if $\lambda_{K_{\PF}} \geq \eta$ but $\lambda_{K_{\PF} + 1}\to 0$ as $M \to \infty$. Then $\bm{A}$ in Theorem~\ref{theorem:Lhat} satisfies
\begin{align*}
    \bm{A} = \begin{cases}
    \diag(a_1,\ldots,a_K) & \text{if $K_{\PF}=0$}\\
    \bm{A}_{\PF}\oplus \diag(a_{K_{\PF}+1},\ldots,a_K) & \text{if $K_{\PF}\in[K-1]$}\\
    \bm{A} & \text{if $K_{\PF}=K$}
    \end{cases}, \quad \norm*{  \bm{A}_{\PF} - \diag(a_1,\ldots,a_{K_{\PF}}) }_2 = O_P(N^{-1/2})
\end{align*}
and, for any fixed set $\mathcal{M} \subset [M]$ and non-pervasive factor $k \in \{K_{\PF}+1,\ldots,K\}$,
\begin{align*}
    \mathop{\sup}\limits_{\substack{ R_m \in \mathscr{C}_1 }}\abs{ \Prob\left[ \mathop{\bigcap}\limits_{m \in \mathcal{M}} \{(N^{-1}+S^{-1})\hat{\sigma}_m^2\}^{-1/2}(\hat{\bm{L}}_{m k}- a_k\bm{L}_{m k}) \in R_m \mid \bm{G},\bm{L},\bm{\Delta},\bm{X} \right] - \prod\limits_{m \in \mathcal{M}}\Phi_1(R_m) }\allowbreak = o_P(1).
\end{align*}
\end{corollary}
The expression for $\bm{A}$ implies we need not worry about the abovementioned contamination, which gives rise to the subsequent result stating the asymptotic distributions for non-pervasive factor loadings are, up to sign, centered around their true loadings. While this is not necessarily true for pervasive factor loadings, it has a negligible effect on inference because contamination will only trivially impact their large non-zero entries and, as illustrated in our data example, pervasive factors are more likely to reflect non-genetic variation. We lastly consider steps~\ref{HiGSS:Delta} and \ref{HiGSS:Post} of Algorithm~\ref{algorithm:HIGSS}.

\begin{theorem}
\label{theorem:Delta}
Suppose Assumptions~\ref{assumption:X}, \ref{assumption:GDelta}, and \ref{assumption:L} hold and $a_1,\ldots,a_K$, $\bm{A}$ are as given in Theorem~\ref{theorem:Ghat} and \ref{theorem:Lhat}. Then the following hold as $N,M,S \to \infty$ for any fixed sets $\mathcal{S} \subset [S]$, $\mathcal{M} \subset [M]$:
\begin{enumerate}[label=(\roman*)]
\item $\mathop{\sup}\limits_{\substack{ R_{sm} \in \mathscr{C}_1 }}\abs{ \Prob\left\{ \mathop{\bigcap}\limits_{\substack{s \in \mathcal{S} \\ m \in \mathcal{M}}} \hat{\sigma}_m^{-1}(\hat{\bm{\Delta}}_{sm}-\bm{D}_{ss}^{1/2}\bm{\Delta}_{sm}) \in R_{sm} \mid \bm{G},\bm{L},\bm{\Delta},\bm{X} \right\} - \prod\limits_{\substack{s \in \mathcal{S} \\ m \in \mathcal{M}}}\Phi_1(R_{sm}) }\allowbreak = \allowbreak o_P(1)$.\label{theorem:Delta:Deltahat}
\item Define $\bm{Z}_s^{\factor} = \hat{\bm{G}}_{s*} - \bm{D}_{ss}^{1/2}\diag(a_1,\ldots,a_K)\bm{G}_{s*}$, $\bm{Z}_m^{\load} = \{(N^{-1}+S^{-1})\hat{\sigma}_m^2\}^{-1/2}(\hat{\bm{L}}_{m*} - \bm{A}\bm{L}_{m*})$, and $z_{sm}^{(\Delta)} = \hat{\sigma}_m^{-1}(\hat{\bm{\Delta}}_{ms} - \bm{D}_{ss}^{1/2}\bm{\Delta}_{ms})$. Then the sets of random variables $\{\bm{Z}_s^{\factor}\}_{s \in \mathcal{S}}$, $\{\bm{Z}_m^{\load}\}_{m \in \mathcal{M}}$, and $\{z_{sm}^{(\Delta)}\}_{(s,m) \in \mathcal{S} \times \mathcal{M}}$ are asymptotically independent conditional on $\{\bm{G},\bm{L},\bm{\Delta},\bm{X}\}$.\label{theorem:Delta:indep}
\end{enumerate}
\end{theorem}
Part~\ref{theorem:Delta:Deltahat} justifies $\hat{\bm{\Delta}}$'s likelihood in Algorithm~\ref{algorithm:HIGSS}, which assumes its entries are independent and normally distributed. The asymptotic independence of the z-statistics in part~\ref{theorem:Delta:indep}, which are the standardized equivalents of $\hat{\bm{G}}$, $\hat{\bm{L}}$, and $\hat{\bm{\Delta}}$, justifies step~\ref{HiGSS:Post}.

\section{Pathway-centric priors and inference in metabolite GWAS}
\label{section:mtGWAS}
Algorithm~\ref{algorithm:HIGSS} provides a general method to compute the posterior of our parameters of interest, where the frequentist estimators in step~\ref{HiGSS:Est}, likelihoods in steps~\ref{HiGSS:Ghat}-\ref{HiGSS:Delta}, and form for the posterior in step~\ref{HiGSS:Post} are consisted across datasets. However, the choice of priors is domain-specific. In this section we describe a set of priors specific to metabolomics, which also highlights the power of our methodology to utilize biologically-informed priors that are intractable in standard Bayesian factor analysis.

\begin{figure}
\centering
\includegraphics[width=0.8\textwidth]{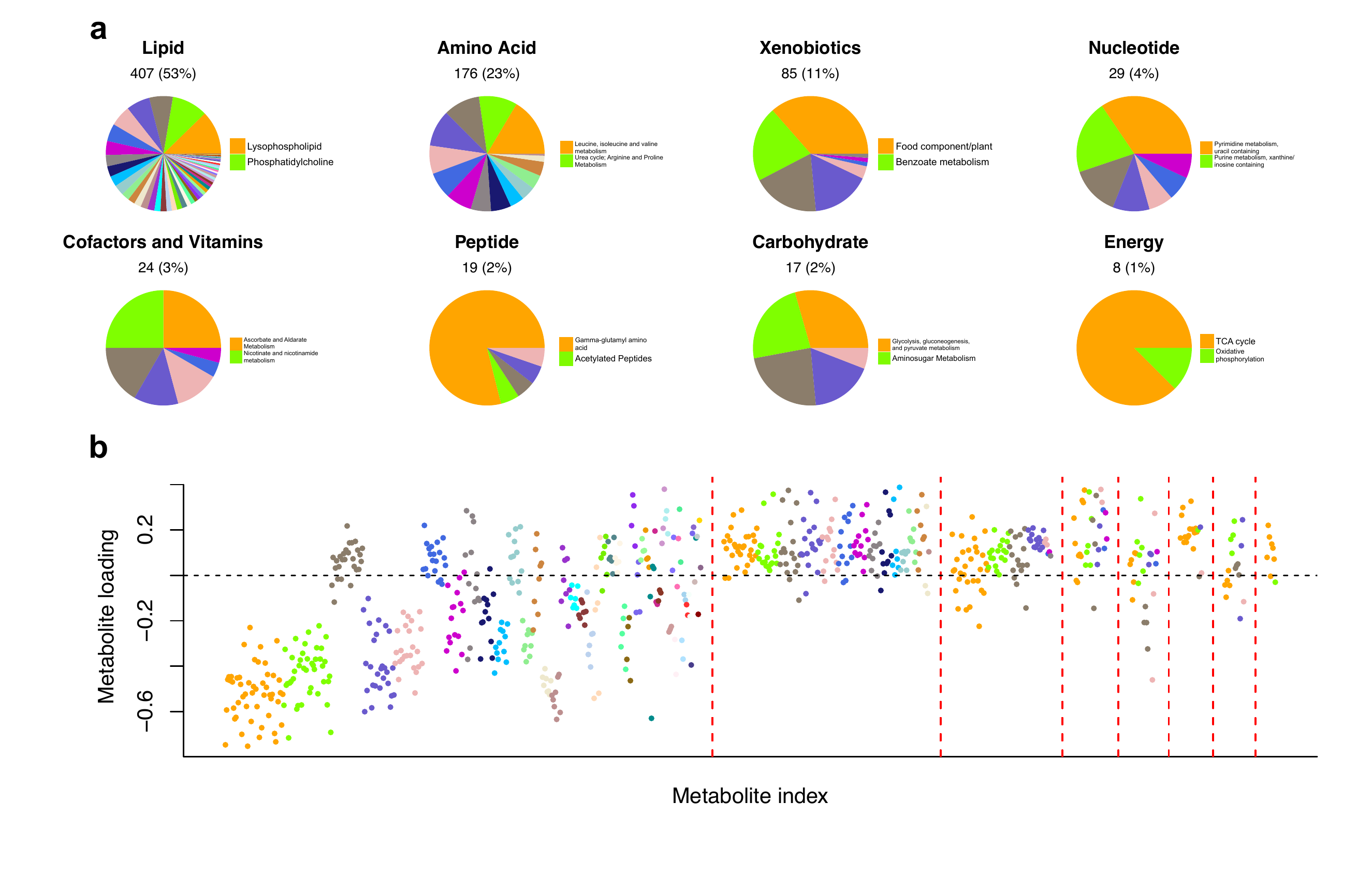}
\caption{(\textbf{a}): Hierarchical partition of metabolites in our data example. Colors indicate each super-pathway's set of sub-pathways and numbers give the number of metabolites in each super-pathway. Each super-pathway's two largest sub-pathways are noted. (\textbf{b}): $\hat{\bm{L}}_{*2}$ in our data example. Red lines demarcate super-pathways and sub-pathway colors match (\textbf{a})'s.}\label{figure:Pathways}
\end{figure}

To motivate our priors, Figure~\ref{figure:Pathways}(a) shows that metabolites can be hierarchically partitioned into a set of super- and sub-pathways, where the overt pathway-specific variation exhibited in Figure~\ref{figure:Pathways}(b) suggests metabolites in the same pathway behave similarly. We therefore follow Section~\ref{subsection:Priors}'s recommendations and design priors for $\bm{L}$ and $\bm{\Delta}$ that incorporate metabolic pathway information.

\subsection{Prior on indirect effects}
\label{subsection:Indirect}
Indirect effects of genotype on metabolite levels are parametrized by $\bm{G}$ and $\bm{L}$. Our prior on $\bm{G}$ follows Section~\ref{subsection:Priors}'s discussion and assumes each column's entries are drawn from a spike and slab:
\begin{align}
\label{equation:Gprior}
    \bm{G}_{sk} \mid (\pi_{G_k}, \tau_{G_k}^2,\bm{D}) \sim (1-\pi_{G_k})\delta_0 + \pi_{G_k}N(0,\tau_{G_k}^2/\bm{D}_{ss}), \quad s \in [S]; k \in [K],
\end{align}
which is conjugate under $\hat{\bm{G}}$'s normal likelihood in Algorithm~\ref{algorithm:HIGSS}. Standardizing by $\bm{D}_{ss}$ implies standardized effects $\bm{D}_{ss}^{1/2}\bm{G}_{sk}$ are identically distributed across SNPs $s$ and reflects the observation that genetic effects are typically inversely proportional to minor allele frequency \citep{MAF_Effects}. We estimate $\pi_{G_k}, \tau_{G_k}^2$ via empirical Bayes.

Our prior on $\bm{L}$ incorporates metabolic pathway information and is constructed to reflect the following observations from Figure~\ref{figure:Pathways}(b) and inference goals. First, metabolite loadings from the same sub-pathway appear to be drawn from the same distribution, where sub-pathway distributions within a super-pathway are more similar than those from different super-pathways. Second, Figure~\ref{figure:Pathways}(b) suggests many sub-pathways have identical loading distributions. For example, the first two orange and green lipid sub-pathways may have the same distributions. Our prior should therefore facilitate a clustering on sub-pathways, which couples biological processes to beget more interpretable inference. Third, it is expected that some metabolites behave as ``outliers'' because they were incorrectly partitioned or do not follow their sub-pathways' distributions.

\begin{figure}
\centering
\includegraphics[width=0.6\textwidth]{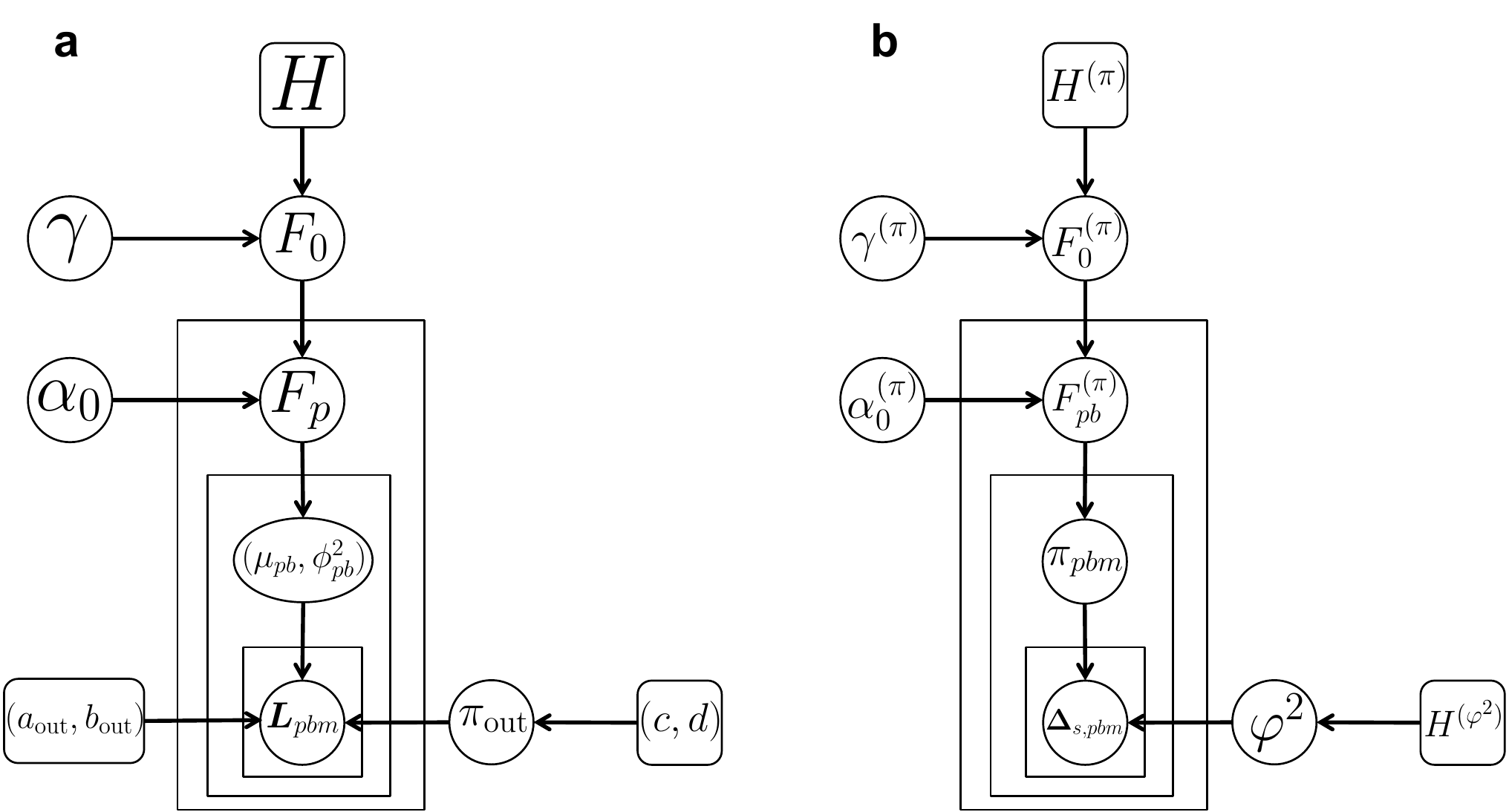}
\caption{A graphical model for $\bm{L}$'s prior (\textbf{a}) and $\bm{\Delta}$'s prior (\textbf{b}). Square and circular nodes are fixed and random quantities. Nodes within a rectangle are replicated.}\label{figure:GraphPrior}
\end{figure}

To accommodate these observations, we use a hierarchical Dirichlet process \citep{HDP} to model sub-pathway distribution parameters. This begets a clustering over sub-pathways, where two sub-pathways from the same super-pathway are \textit{a priori} more likely to lie in the same cluster than sub-pathways from different super-pathways. Metabolite loadings are then either drawn from their sub-pathway's distribution or, to accommodate the above third observation, an ``outlier'' distribution. Mathematically, let $p \in [P]$ index super-pathways, $b \in [B_p]$ index constituent sub-pathways, and $m \in [M_{pb}]$ index constituent metabolites. We assume $\bm{L}$ has independent columns and, to simplify notation, suppress any dependence on the factor number. As such, we let $\bm{L}_{pbm}\in \mathbb{R}$ be a factor's loading for the $m$th metabolite in the $b$th and $p$th sub- and super-pathway for the remainder of Section~\ref{subsection:Indirect}. Then for concentration parameters $\gamma,\alpha_0>0$ and probability measure $H$ on $\mathbb{R} \times \mathbb{R}_{\geq 0}$,
\begin{align}
\label{equation:HDP}
\begin{aligned}
    &F_0 \mid (\gamma, H) \sim \DP(\gamma, H), \quad \pi_{\outlier}\sim \text{Beta}(c,d)\\ &\text{$F_p \mid (\alpha_0, F_0) \sim \DP(\alpha_0, F_0), \quad p \in [P]$}\\
    &\text{$(\mu_{pb}, \phi^2_{pb}) \mid F_p \sim F_p, \quad b \in [B_p]$}\\
    &\text{$\bm{L}_{pbm} \mid (\mu_{pb}, \phi^2_{pb}, \pi_{\outlier},a_{\outlier},b_{\outlier}) \sim (1-\pi_{\outlier})N( \mu_{pb}, \phi^2_{pb} ) + \pi_{\outlier}U[a_{\outlier},b_{\outlier}], \quad m \in [M_{pb}]$}.
\end{aligned}
\end{align}
Figure~\ref{figure:GraphPrior}(a) provides a graphical description of this prior. Assuming for the moment that $\pi_{\outlier}=0$, $\bm{L}_{pbm}$ is drawn from a hierarchical Dirichlet process mixture model, where mixture components (clusters) are defined by the unique values of $\{(\mu_{pb},\phi^2_{pb})\}_{p\in[P];b\in[B_p]}$ \citep{HDP}. Briefly, the probability measure $F_0$ is first drawn from a Dirichlet process with baseline measure $H$ and, with probability one, takes the form $\sum_{j=1}^{\infty} \beta_{0,j} \delta_{(\mu_j,\phi^2_j)}$ for some probability weights $\beta_{0,j}$ and $(\mu_j,\phi^2_j) \in \mathbb{R} \times \mathbb{R}_{\geq 0}$. Each $F_p$ is subsequently drawn from another Dirichlet process with baseline measure $F_0$, meaning $F_p$ can also be expressed as $\sum_{j=1}^{\infty} \beta_{p,j} \delta_{(\mu_j,\phi^2_j)}$ for some probability weights $\beta_{p,j}$. The third line in \eqref{equation:HDP} implies sub-pathways within a super-pathway are assigned to clusters based on their parameters $(\mu_{pb},\phi^2_{pb})$, where sub-pathways that share parameters fall in the same cluster. However, since the $F_p$'s share atoms, sub-pathways from different super-pathways can be assigned to the same cluster, where larger values of $\alpha_0$ increase the probability of this occurring.

The parameter $\pi_{\outlier}$ in \eqref{equation:HDP} is the fraction of all metabolites that are ``outliers'', i.e. metabolites whose loadings do not follow the hierarchical Dirichlet process mixture model. Since we assume their loadings can take any value with equal probability, we treat them as uniformly distributed and set $a_{\outlier},b_{\outlier}$ so $[a_{\outlier},b_{\outlier}]$ contains all observed loadings. We let $\pi_{\outlier} \sim \text{Beta}(c,d)$ so as to allow the data to help dictate outlier frequency, and set $c=1/5,d=999/5$ so that out of 1,000 metabolites, we \textit{a priori} expect one outlier and no more than five outliers with probability 0.95.

Similar to \citet{HDP}, we place diffuse priors on $\gamma$ and $\alpha_0$. We set the base distribution $H$ so that $\Prob\{(\mu_{pb},\phi_{pb}^2) = (0,0)\mid H\} > 0$ and $\mu_{pb} \indep \phi_{pb}^2\mid \{H, (\mu_{pb},\phi_{pb}^2) \neq (0,0)\}$. The former reflects our belief that some sub-pathways will have loadings exactly equal to 0 and we let $\mu_{pb} \mid \{H, (\mu_{pb},\phi_{pb}^2) \neq (0,0)\}$ follow a spike and mixture-normal slab to ensure conjugacy, where the slab is chosen to approximate the horseshoe prior \citep{Horseshoe}. We let $\phi_{pb}^2\mid \{H, (\mu_{pb},\phi_{pb}^2) \neq (0,0)\}$ be a densely discretized gamma distribution, where discretizing facilitates faster computation by letting us to integrate out $\bm{L}$ in our Gibbs sampler. Section~\ref{appendix:section:GibbsSampler} provides additional details and describes our Gibbs sampler, which utilizes the  Chinese restaurant franchise parametrization of the hierarchical Dirichlet process to efficiently sample from the posterior.



\subsection{Prior on direct effects}
\label{subsection:Direct}
Similar to our prior on $\bm{L}$, we utilize a hierarchical Dirichlet process mixture model to place a prior on $\bm{\Delta}$'s hyperparameters in Algorithm~\ref{algorithm:HIGSS}. However, it differs from the prior presented Section~\ref{subsection:Indirect} in two important ways. First, it only considers sub-pathways and ignores super-pathway labels. While this does discard information, we found that incorporating super-pathways led to trivial changes in estimates and was not worth the extra computation. Second, our model for $\bm{\Delta}$'s hyperparameters induces a clustering on metabolites as opposed to sub-pathways, which helps pool information across related metabolites to improve otherwise uncertain hyperparameter estimates.

We build $\bm{\Delta}$'s prior to resemble that in \citet{Biostat_metab}, which modeled genetic effects on metabolites as a spike and normal-slab with metabolite-specific spike probabilities and constant slab variance. We use the same super- and sub-pathway notation to be consistent with Section~\ref{subsection:Indirect} and re-index $\bm{\Delta}$ for the purposes of Section~\ref{subsection:Direct} so that $\bm{\Delta}_{s,pbm}$ is the direct effect of the $s$th SNP on metabolite $m\in[M_{pb}]$ in sub-pathway $b\in[B_p]$ and super-pathway $p \in [P]$. Then
\begin{align}
\label{equation:HDP:direct}
\begin{aligned}
    & F_0^{\direct} \mid \{\gamma^{\direct},H^{\direct}\} \sim \DP(\gamma^{\direct},H^{\direct}), \quad \varphi^2\mid H^{\directvar} \sim  H^{\directvar}\\
    &\text{$F_{pb}^{\direct} \mid \{\alpha_0^{\direct},F_0^{\direct}\} \sim  \DP(\alpha_0^{\direct},F_0^{\direct}), \quad p \in [P]; b \in [B_p]$}\\
    &\text{$\pi_{pbm} \mid F_{pb}^{\direct} \sim  F_{pb}^{\direct}, \quad m \in [M_{pb}]$}\\
    &\text{$\bm{\Delta}_{s,pbm} \mid (\pi_{pbm}, \varphi^2,\bm{D}) \sim  (1-\pi_{pbm})\delta_0 + \pi_{pbm}N\left(0, \varphi^2 \sigma_{pbm}^2/\bm{D}_{ss} \right), \quad s\in[S],$}
\end{aligned}
\end{align}
where $H^{(\pi)}$ and $H^{(\varphi^2)}$ are probability measures on $[0,1]$ and $(0,\infty)$ and $\sigma_{pbm}^2$ is the re-indexed version of $\sigma_m^2$ defined in \eqref{equation:Model}. We replace it with its consistent estimator defined in step~\ref{HiGSS:Lhat} of Algorithm~\ref{algorithm:HIGSS} in practice. Standardizing $\bm{\Delta}$'s slab variance by $\sigma_{pbm}^2$ adjusts for heterogeneity in metabolite levels, and dividing by $\bm{D}_{ss}$ reflects the fact that genetic effects are often inversely proportional to minor allele frequency. Figure~\ref{figure:GraphPrior}(b) gives a graphical depiction of \eqref{equation:HDP:direct} where, just like we did in Section~\ref{subsection:Indirect}, we place diffuse priors on the concentration parameters $\gamma^{(\pi)},\alpha_0^{(\pi)}$.

The observed data $\hat{\bm{\Delta}}$ will be informative for $\varphi^2$ because it is shared across all SNPs and metabolites. However, $\pi_{pbm}$ is metabolite-specific which, given $\bm{\Delta}$'s sparsity, suggests its estimates may be uncertain. To address this, our hierarchical Dirichlet process mixture prior on $\pi_{pbm}$ induces a clustering on metabolites, where metabolites in the same cluster share $\pi_{pbm}$'s and are more likely to be grouped with other metabolites in the same sub-pathway. This pools information across related metabolites, thereby reducing our uncertainty in $\pi_{pbm}$.

It remains to specify $H^{(\pi)}$ and $H^{(\varphi^2)}$, where care must be taken to ensure inference is computationally tractable. Let $\Theta= \{\{\pi_{pbm}\}_{p \in [P]; b\in[B_p]; m \in [M_{pb}]}, \varphi^2\}$. Then $\bm{\Delta}$'s posterior is
\begin{align}
\label{equation:PostDelta}
    \Bprob\{\bm{\Delta} \mid \hat{\bm{\Delta}},\bm{D},H^{(\pi)},H^{(\varphi^2)}\} =& \smallint \Bprob( \bm{\Delta} \mid \hat{\bm{\Delta}},\bm{D}, \Theta ) \Bprob\{\Theta \mid \hat{\bm{\Delta}},\bm{D}, H^{(\pi)},H^{(\varphi^2)}\}d \Theta.
\end{align}
The first posterior in the integral matches that in step~\ref{HiGSS:Delta} of Algorithm~\ref{algorithm:HIGSS} and is an entry-wise spike and normal-slab distribution with a closed form. However, $\Theta$'s posterior is generally intractable, meaning the integral must be approximated via Markov chain Monte Carlo. Unfortunately, a Gibbs sampler with standard conjugate inverse-gamma and beta distributions for $H^{(\varphi^2)}$ and $H^{(\pi)}$ would require sampling $\bm{\Delta}$ and would result in intractably slow mixing due $\bm{\Delta}$'s high dimension. We address this by letting $H^{(\pi)}$ and $H^{(\varphi^2)}$ be discrete. Briefly, let $\{0\} \cup \mathcal{X}^{(\pi)} \subset [0,1)$ and $\mathcal{X}^{(\varphi^2)} \subset (0,\infty)$ be finite sets containing all possible values of $\pi_{pbm}$ and $\varphi^2$. We then set $H^{(\pi)}$ and $H^{(\varphi^2)}$ so that
\begin{align*}
    &\Prob\{\pi_{pbm}=0 \mid H^{(\pi)}\} = 0.5, \quad \text{$\Prob\{\pi_{pbm}=x \mid H^{(\pi)}\} = 0.5/\abs*{\mathcal{X}^{(\pi)}}$ for all $x \in \mathcal{X}^{(\pi)}$}\\
    &\text{$\Prob\{\varphi^2 = x \mid H^{(\varphi^2)}\} = 1/\abs*{ \mathcal{X}^{(\varphi^2)} }$ for all $x \in \mathcal{X}^{(\varphi^2)}$},
\end{align*}
where $H^{(\pi)}$'s non-trivial mass at 0 reflects our expectation that some metabolites will have no direct effects. Our software-default of 0.5 is likely conservative given that the levels of many metabolites are impacted by at least one SNP \citep{mtGWAS_pleiotropy}. Critically, the above discrete distribution facilitates faster mixing in our Gibbs sampler to approximate \eqref{equation:PostDelta} by allowing us to integrate out $\bm{\Delta}$, and lets us pre-compute the likelihood $\Bprob\{\hat{\bm{\Delta}} \mid \Theta, \bm{D},H^{(\pi)},H^{(\varphi^2)}\} = \Bprob(\hat{\bm{\Delta}} \mid \Theta, \bm{D}) $ for all $\Theta$ required to propose new states, which significantly reduces computation time. Section~\ref{appendix:section:DirectEffects} details how we set $\mathcal{X}^{(\pi)}$ and $\mathcal{X}^{(\varphi^2)}$.

\subsection{Pathway-specific inference on direct effects}
\label{subsection:DirectPathway}
Our prior on direct effects in \eqref{equation:HDP:direct} lets us compare the ``direct'' heritability of metabolic sub-pathways. To see this, let $\Theta$ be as defined in Section~\ref{subsection:Direct} and $\bm{Y}_{i,pbm}$ be individual $i$'s abundance of metabolite $m$ in sub- and super-pathways $b$ and $p$. Then under model~\eqref{equation:Model} and $\bm{\Delta}$'s prior in \eqref{equation:HDP:direct}, the heritability  of metabolite $m$ conditional on latent factors $\bm{C}$ is
\begin{align*}
    h_{pbm}^2 := \frac{ N^{-1}\sum_{i=1}^N \V(\textstyle\sum_{s=1}^S \bm{X}_{is} \bm{\Delta}_{s,pbm} \mid \bm{L},\bm{C},\Theta) }{ N^{-1}\sum_{i=1}^N \V(\bm{Y}_{i,pbm}  \mid \bm{L},\bm{C},\Theta) } = \frac{\varphi^2\pi_{pbm}}{\varphi^2\pi_{pbm} + 1} ,
\end{align*}
where the numerator and denominator in the first fraction are the average variance of the metabolite's direct genetic effect and abundance, respectively. Since a metabolite's heritability is an increasing function of $\pi_{pbm}$, evaluating sub-pathway $b$ in super-pathway $p$'s heritability is equivalent to studying the distribution of $\{\pi_{pbm}\}_{m \in [M_{pb}]}$. This distribution is exactly $F_{pb}^{(\pi)}$ defined in \eqref{equation:HDP:direct}, where a sub-pathway is more heritable when draws from $F_{pb}^{(\pi)}$ tend to be large. We therefore define sub-pathway $b$ in super-pathway $p$'s direct heritability score, $DHS_{pb}$, to be
\begin{align}
\label{equation:DHS}
    DHS_{pb} = \Prob\{ f_{pb}^{(\pi)} > f_0^{(\pi)} \mid \hat{\bm{\Delta}},\bm{D},H^{(\pi)},H^{(\varphi^2)}\}, \quad f_{pb}^{(\pi)} \mid F_{pb}^{(\pi)} \sim F_{pb}^{(\pi)}, \quad f_{0}^{(\pi)} \mid F_{0}^{(\pi)} \sim F_{0}^{(\pi)}.
\end{align}
Since $F_{0}^{(\pi)} = \E\{ F_{pb}^{(\pi)} \mid  F_{0}^{(\pi)}\}$, $DHS_{pb}$ is interpretable as the probability a randomly chosen metabolite from sub-pathway $b$ is more heritable than one from the average sub-pathway. These scores allow us to rank sub-pathways in terms of their heritability, where sub-pathways with larger $DHS_{pb}$ are more heritable. We demonstrate their utility in  Section~\ref{section:RealData}.

\section{Simulations}
\label{section:Simulations}

\subsection{Simulation setup}
\label{subsection:SimulationSetup}
We simulated metabolomic data from $N=500$ individuals  whose $M=257$ metabolites were regulated by $S=5,000$ SNPs to evaluate our estimator for the number of latent factors and Algorithm~\ref{algorithm:HIGSS}. Unobserved genotypes $\bm{X}$ and metabolite levels $\bm{Y}$ were generated  assuming $K=10$ latent factors:
\begin{align}
\label{equation:Simulations}
\begin{aligned}
    &f_s \sim U[0.05,0.5], \quad \bm{X}_{is} \sim \text{Bin}(2,f_s),\quad i \in [N]; s \in [S]\\
    & \bm{G}_{sk} \sim (1-10^{-3})\delta_0 + 10^{-3} N(0,2/N), \quad \bm{L}_{*k} \sim HDP(\lambda_k),\quad s \in [S]; k \in [K]\\
    &\bm{\Delta}_{sm} \sim (1-\pi_{\Delta_m})\delta_0 + \pi_{\Delta_m} N(0,\tau_{\Delta s}^2), \quad \sigma_m^2 \sim \text{Gamma}(1,1), \quad s \in [S]; m \in [M]\\
    & \bm{C} \sim MN(\bm{X}\bm{G}, I_N, I_K), \quad \bm{Y} \sim MN( \bm{C}\bm{L}^{\top} + \bm{X}\bm{\Delta}, \diag(\sigma_1^2,\ldots,\sigma_M^2), I_N ).
\end{aligned}
\end{align}
For each dataset we defined GWAS summary statistics $\hat{\beta}_{sm}$ to be estimates for the slope in the regression of $\bm{Y}_{*m}$ onto $\bm{X}_{*s}$ and the observed matrix $\hat{\bm{B}}$ such that $\hat{\bm{B}}_{sm} = (\bm{X}_{*s}^{\top}\bm{P}_{\bm{1}_N}^{\perp}\bm{X}_{*s})^{1/2}\hat{\beta}_{sm}$ for $\bm{P}_{\bm{1}_N}^{\perp}$ the orthogonal projection matrix that projects vectors onto the kernel of $(1,\ldots,1)\in\mathbb{R}^{1 \times N}$.

The parameter $\pi_{\Delta_m}$ was simulated as $\pi_{\Delta_m}\sim 0.3\delta_0+0.7 F$ for $F$ a truncated exponential distribution between 0 and 0.01 with mean $0.005$, and $\tau^2_{\Delta s}=5^2/(\bm{X}_{*s}^{\top}\bm{P}_{\bm{1}_N}^{\perp}\bm{X}_{*s})$. These, along with $\bm{G}$'s hyperparameters, were all commensurate to the values estimated in our real data example. We drew $\bm{L}$'s columns independently from a hierarchical Dirichlet process (HDP) mixture model by partitioning metabolites into super- and sub-pathways such that the number of constituent metabolites in each partition mirrored our real data example (see Section~\ref{appendix:section:SimL} for details). The parameters $\lambda_k=M^{-1}\E(\bm{L}_{*k}^{\top} \bm{L}_{*k})$ dictate the strength of each latent factor where, as we discussed in Section~\ref{subsection:theory}, factors with larger $\lambda_k$ are stronger and easier to recover. To make simulations realistic, we let $\lambda_1 > \cdots >\lambda_K$ take a wide range of values, where $\lambda_1=3.35$ corresponded to a large pervasive factor and $\lambda_K=0.25$  was weaker. Notably, these values implied there was a small eigengap between the signal and noise, since the $K$th singular value of $\hat{\bm{B}}$ was on the average only 14.5\% larger than $\norm*{\hat{\bm{B}} - \E(\hat{\bm{B}} \mid \bm{X},\bm{C},\bm{L},\bm{\Delta})}_2$, the size of the noise. Consequently, as is the case in real data, recovering all factors was non-trivial.

\subsection{Simulation results}
\label{subsection:SimulationResults}
\begin{figure}
\centering
\includegraphics[width=0.9\textwidth]{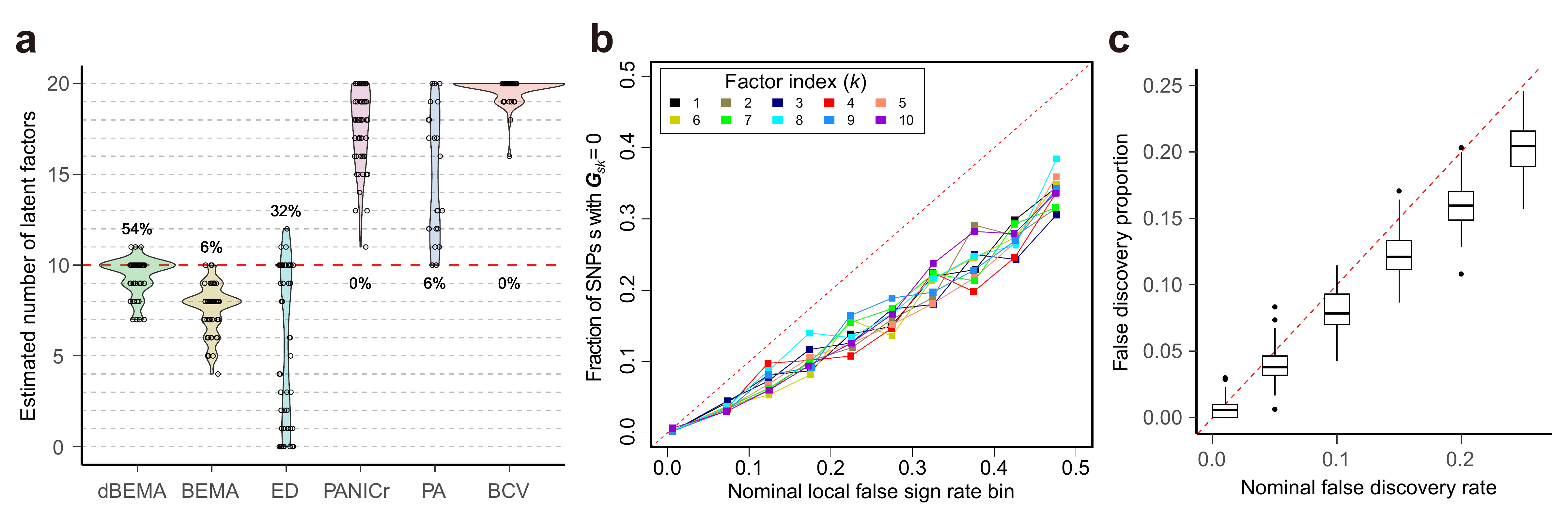}
\caption{Simulation results for 50 simulated datasets. (\textbf{a}): Estimates for $K$. dBEMA is our proposed method and BCV's maximum $K$ was set to 20. Percents give the fraction of times the estimate matched the true $K=10$. (\textbf{b}): Inference on $\bm{G}$. Local false sign rates were binned into 10 bins. (\textbf{c}): Inference on $\bm{\Delta}$. Each point represents a simulated dataset.}\label{Figure:simulations}
\end{figure}

We first considered estimating the number of latent factors $K$. We compared our method described in Section~\ref{subsection:K} (dBEMA) to bulk eigenvalue matching (BEMA) proposed in \citet{ke2021estimation}, the eigenvalue difference approach (ED) taken in \citet{onatski2010determining}, the information criterion (PANICr) from \citet{bai2002determining}, parallel analysis (PA) \citep{SVA}, and bi-cross validation (BCV) \citep{owen2016bi}. Figure~\ref{Figure:simulations}(a) contains the results, where our method clearly outperforms existing methods. Notably, PA and BCV, which are ubiquitous in omic analysis pipelines, severely overestimate $K$. This is because the error matrix $\tilde{\bm{E}}$ in the expression for $\hat{\bm{B}}$ in \eqref{equation:Bhat} has dependent rows (SNPs), which belies their assumption that the error matrix has independent entries. As a consequence, they mistake the error's row-wise dependence for latent structure.

Assuming $K$ was known, we next considered estimating $\bm{G}$ and $\bm{L}$, which parameterize indirect effects, and direct effects $\bm{\Delta}$. We first used Algorithm~\ref{algorithm:HIGSS} and the priors described in Sections~\ref{subsection:Indirect} to estimate $\bm{G}$ and $\bm{L}$. To address issues of factor identifiability and match estimated and simulated factors, we rotated our estimators $\hat{\bm{G}}$ and $\hat{\bm{L}}$ defined in Algorithm~\ref{algorithm:HIGSS} by solving Procrustes problem $\text{argmin}_{\bm{Q}^{\top}\bm{Q}=I_K} \norm*{ \hat{\bm{G}}\bm{Q} - \tilde{\bm{G}} }_F$ for $\tilde{\bm{G}}$ defined in \eqref{equation:Bhat}. We used the local false sign rate ($\lfsr$) to perform inference on $\bm{G}$:
\begin{align}
\label{equation:lfsr}
    \lfsr(\bm{G}_{sk}) = 1-\max\{ \Prob(\bm{G}_{sk} > 0 \mid \hat{\bm{B}},\bm{D}), \Prob( \bm{G}_{sk} < 0 \mid \hat{\bm{B}},\bm{D} ) \}, \quad s \in [S]; k \in [K].
\end{align}
The $\lfsr$ is the probability a coefficient's inferred non-zero sign is incorrect and is at least as large as the local false discovery rate. It is used here to avoid issues with the local false discovery rate that can arise when the null and alternative distributions are hard to distinguish \citep{stephens2017false}. Figure~\ref{Figure:simulations}(b) shows that our estimate for $\bm{G}$'s $\lfsr$ controls the local false discovery rate. Section~\ref{appendix:section:SimL} contains results for $\bm{L}$, which illustrate the importance of our novel theoretical results from Section~\ref{subsection:theory} giving $\hat{\bm{L}}$'s asymptotic distribution.

We lastly estimated $\bm{\Delta}$. Since the prior for $\bm{\Delta}$ recommended for metabolomic data in Section~\ref{subsection:Direct} requires non-trivial computation, we perform inference in these simulated data by controlling the false discovery rate via the Benjanmini-Hochberg procedure \citep{BH} after using $\hat{\bm{\Delta}}$'s likelihood in Algorithm~\ref{algorithm:HIGSS} to compute p-values for the null hypotheses $H_{0,sm}: \bm{\Delta}_{sm}=0$. Figure~\ref{Figure:simulations}(c) shows this controls the false discovery rate, indicating $\hat{\bm{\Delta}}$'s likelihood in Algorithm~\ref{algorithm:HIGSS} is appropriate.

\section{Real data analysis}
\label{section:RealData}
We demonstrate the power of our methodology using metabolite summary statistics derived from $N=6,136$ Finnish adults participating in the METSIM study \citep{Metsim}. Briefly, we considered SNPs with minor allele frequencies $\geq 5\%$ and subsequently pruned them for linkage disequilibrium. Metabolites without a name or missing in more than 20\% of samples were excluded, which resulted in $S=70,140$ SNPs and $M=765$ metabolites. Note that while some metabolites had missing data, the missingness was likely inconsequential, as more that 80\% of metabolites had less than 5\% missing data. Figure~\ref{figure:Pathways}(a) provides an overview of the hierarchical partition of metabolites into super- and sub-pathways.

\subsection{Indirect effects}
\label{subsection:RealData:Indirect}

We first estimated the number of latent factors $K$ and indirect effects. Table~\ref{table:IndirectOverview}(a) gives each method's estimate for $K$, where our method dBEMA estimates 36. Compared to dBEMA, the methods ED, PANICr, and BCV behave as they did in simulations, where ED likely underestimates and PANICr and BCV likely overestimate $K$. We then used Algorithm~\ref{algorithm:HIGSS} with $\hat{K}=36$ latent factors to determine $\bm{G}$ and $\bm{L}$'s posterior using Section~\ref{subsection:Indirect}'s priors, and determined the SNPs and metabolites involved in indirect effects using the local false sign rate (lfsr) defined in \eqref{equation:lfsr}, where $\lfsr(\bm{L}_{mk})$ was defined by replacing $\bm{G}_{sk}$ with $\bm{L}_{mk}$ in \eqref{equation:lfsr}. We defined a ``genetic factor'' to be a factor $k$ satisfying $\lfsr(\bm{G}_{sk}) \leq 0.05$ for at least one SNP $s \in [S]$, an ``indirect SNP'' to be a SNP $s$ with $\lfsr(\bm{G}_{sk}) \leq 0.05$ for at least one $k \in [\hat{K}]$, and an ``indirect metabolite'' to be a metabolite $m$ satisfying $\lfsr(\bm{L}_{mk}) \leq 0.05$ for at least one genetic factor $k$. Table~\ref{table:IndirectOverview}(b) provides the results, where all but factors 1, 2, 3, and 6 are genetic factors. As these factors have the largest loadings, it suggests large, pervasive factors reflect uninteresting variation and reinforces the importance of non-pervasive factors.

\begin{table}
\centering
\includegraphics[width=0.6\textwidth]{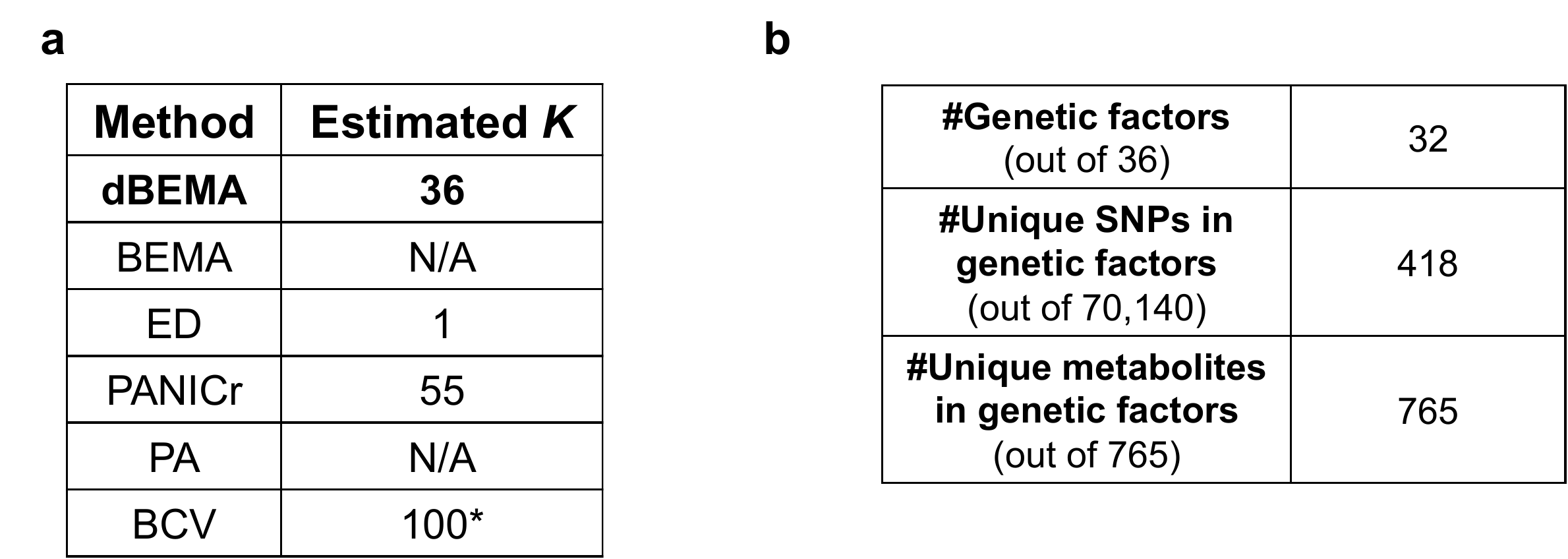}
\caption{(\textbf{a}): Estimated number of factors. ``N/A'' means the method's software could not be run on these data; the maximum number of factors considered by BCV was 100. (\textbf{b}): Indirect effect results.  }\label{table:IndirectOverview}
\end{table}

Table~\ref{table:IndirectOverview}(b) implies the levels of all metabolites are regulated by a small number of SNPs, which suggests these SNPs are highly pleiotropic. Our results are substantiated by \citet{mtGWAS_pleiotropy}, who studied 233 serum metabolites measured in 136,016 individuals and found that all metabolites were regulated by 276 genomic regions. Remarkably, these regions contained 306 (73\%) of the indirect SNPs we identified, which is more than four times more than expected by chance alone (p-value $<10^{-16}$). 

\begin{figure}
\centering
\includegraphics[width=1\textwidth]{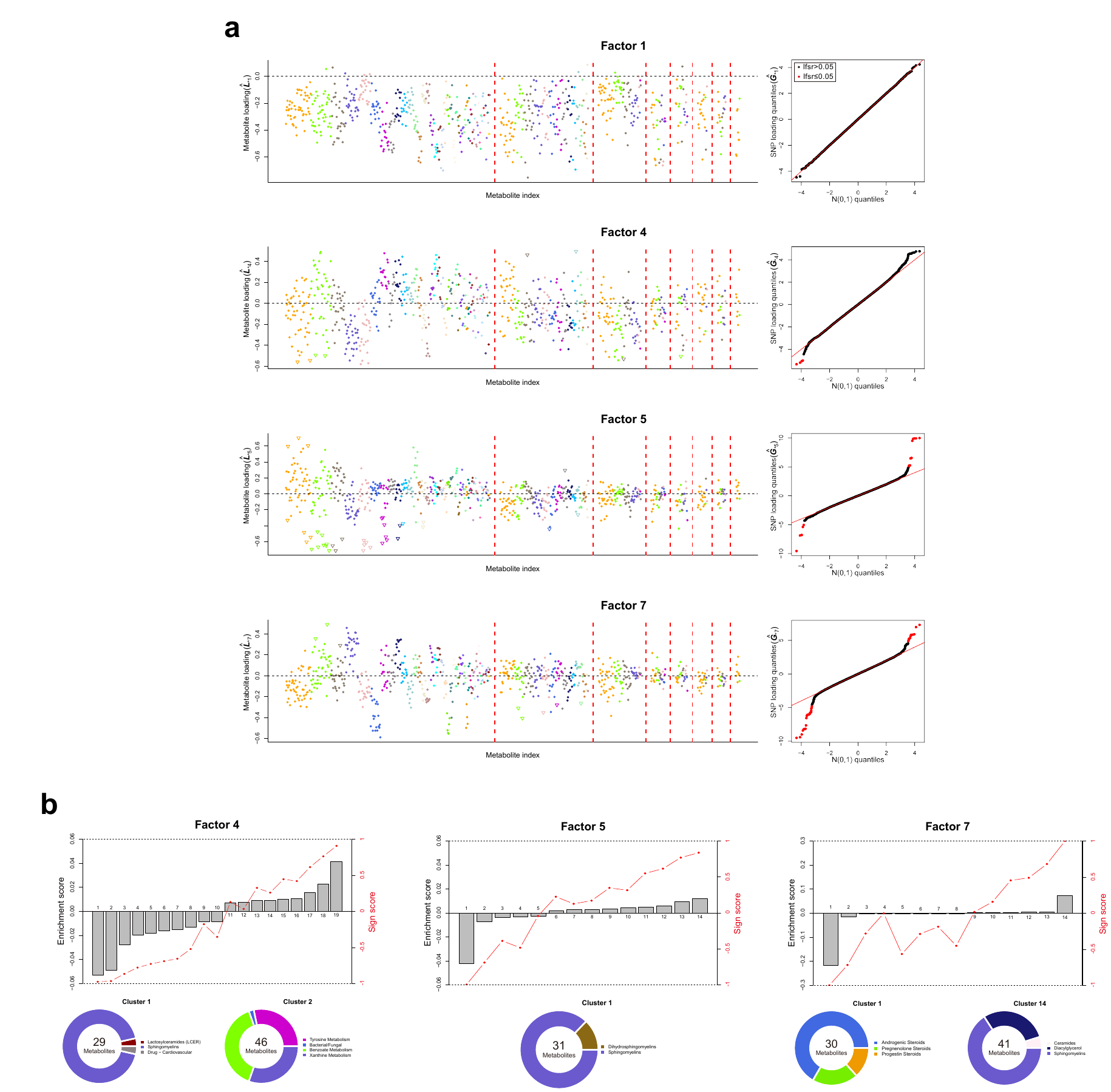}
\caption{(\textbf{a}): Metabolite and SNP loadings for factor 1 and the first three genetic factors. Dashed red lines demarcate super-pathways whose names are given in Figure~\ref{figure:Pathways}(a) and each set of colored metabolites is a sub-pathway; `$\triangledown$' points indicate outlier metabolites in that factor, defined as metabolites whose posterior probability of being an outlier was $\geq 0.95$. (\textbf{b}): Enrichment and sign scores for clusters identified in the three genetic factors, where the columns of $\hat{\bm{L}}$ were scaled to all have the same norm prior to running the Gibbs sampler to make scores comparable across factors. Scores are negative if more of the cluster's estimated loadings are negative than positive.}\label{Figure:IndirectPathway}
\end{figure}

Since all metabolites appear to be indirectly genetically regulated, simply reporting indirect metabolites will do nothing to aid biological interpretation. Instead, we interpret factors mediating genetic effects by studying how their loadings behave across metabolic super- and sub-pathways using $\bm{L}$'s posterior.  As discussed in Section~\ref{subsection:Indirect}, $\bm{L}$'s posterior begets a clustering on sub-pathways that couples biological processes to improve inference. We illustrate the utility of this approach using the first three genetic factors whose estimated metabolite and SNP loadings are plotted in the last three rows of Figure~\ref{Figure:IndirectPathway}(a). For each factor $k \in [\hat{K}]$, we used our Gibbs sampler detailed in Section~\ref{appendix:section:GibbsSampler} to sample from $\Bprob[ \{(\mu_{pb},\phi_{pb}^2)\}_{p\in[P];b\in[B_p]} \mid \hat{\bm{B}},\bm{D} ]$, where parameters $(\mu_{pb},\phi_{pb}^2)$ are defined in \eqref{equation:HDP} and implicitly depend on factor $k$. These parameters characterize the distribution of factor $k$'s loadings for metabolites in sub-pathway $b$ and super-pathway $p$, where a ``cluster'' is a set of sub-pathways that share parameters. We defined the maximum \textit{a posteriori} clustering to be our hard clustering, which we found gave the most parsimonious results compared to other hard clustering methods. We determined a cluster $\mathcal{C}$'s importance with enrichment ($\mathscr{E}$) and sign ($\mathscr{S}$) scores:
\begin{align*}
    \mathscr{E}_{\mathcal{C}} = \abs*{\mathcal{C}}^{-1} \textstyle \sum_{(p,b) \in \mathcal{C}} \E( \mu_{pb}^2 + \phi_{pb}^2 \mid \hat{\bm{B}},\bm{D} ), \quad \mathscr{S}_{\mathcal{C}} =& \max  \textstyle[ 2\abs*{\mathcal{C}}^{-1}\sum_{(p,b) \in \mathcal{C}} \Prob\{ N(\mu_{pb},\phi_{pb}^2) > 0\mid \hat{\bm{B}},\bm{D} \}\\ & \textstyle 2\abs*{\mathcal{C}}^{-1}\sum_{(p,b) \in \mathcal{C}} \Prob\{ N(\mu_{pb},\phi_{pb}^2) < 0\mid \hat{\bm{B}},\bm{D} \} ]-1.
\end{align*}
The enrichment score is the average squared magnitude of the cluster's loadings. The sign score is bounded between 0 and 1 and will be close to 1 if the cluster's loading distribution is far away from zero and 0 if it is symmetric around zero. Clusters with large scores have large loadings with the same non-zero signs, and are therefore ideal candidates to help infer a factor's function.



Figure~\ref{Figure:IndirectPathway}(b) gives a compendious overview of the clustering results for the first three genetic factors, where the pie charts give memberships for clusters with the largest enrichment scores. We only display one cluster for factor 5 because its second highest enrichment and sign scores were small. Remarkably, our unsupervised clusters mirror biology. Cluster 1 in factors 5 and 7 contain both sphingomyelin sub-pathways and all three classes of sex hormones, respectively. Grouping sphingomyelins with ceramides in factors 4 and 7 is expected given that a ubiquitous reaction produces sphingomyelins from ceramides, where the inclusion of diacylglycerols in cluster 14 is likely because the reaction creates them as a by-product \citep{SphingoCeramide}.

We use these clusters and their corresponding significant SNPs (the red dots in Figure~\ref{Figure:IndirectPathway}(b)) to infer indirect mechanisms by which genotype systematically regulates metabolite levels.

\begin{figure}
\centering
\includegraphics[width=0.9\textwidth]{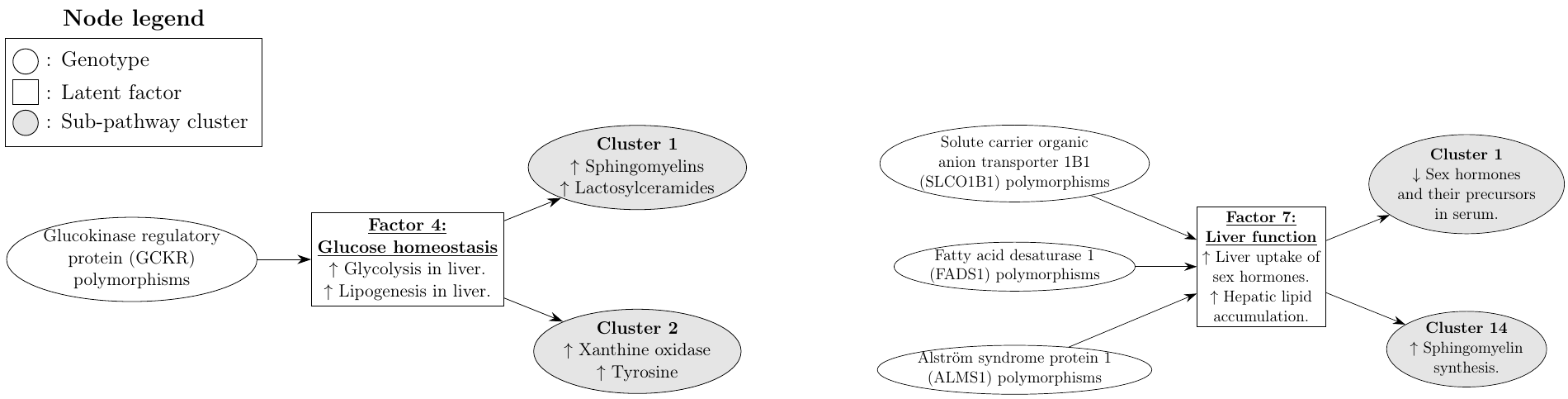}
\caption{Interpretations of factors 4 and 7.}\label{Figure:FactorInt}
\end{figure}

\vspace{1mm}
\noindent \textbf{Factor 4: Glucose homeostasis.} Factor 4 contains four significant SNPs that all map to the gene GCKR, which binds to and inactivates the enzyme glucokinase in the liver to inhibit hepatic glycolysis and subsequent lipogenesis \citep{GCKR}. Since the four SNPs' minor alleles reduce GCKR's ability to bind to glucokinase \citep{GCKR}, we posit factor 4 relates to glucose homeostasis characterized by an up-regulation of glycolysis and lipogenesis in the liver (Figure~\ref{Figure:FactorInt}). Excess glycolysis has the effect of increasing the levels of tyrosine and its derivatives, which include microbiome-derived benzoates \citep{MicrobeTyrosine}, as well as possibly increasing the activity of the enzyme xanthine oxidase that is responsible for many metabolites in the xanthine sub-pathway \citep{Tyrosine,Xanthine}. Hepatic lipogenesis raises the levels of sphingomyelin and lactosylceramide metabolites \citep{SphingoPathway}.

\vspace{1mm}
\noindent \textbf{Factor 7: Liver function.} 43 of the 54 SNPs regulating factor 7 could be mapped to one of the three genes in Figure~\ref{Figure:FactorInt}. SLCO1B1 is uniquely expressed in the liver and encodes the uptake transport protein responsible for transporting sex hormones and their related metabolites \citep{SLCO_1}. The SLCO1B1 polymorphisms we identified tend to increase its activity \citep{SLCO_2}, which likely explains the decrease in serum sex hormone levels. FADS1 encodes the rate-limiting enzyme in the synthesis of long chain polyunsaturated fatty acids (LC-PUFAs), where the polymorphisms we identify reduce its expression to beget hepatic lipid accumulation and subsequent sphingomyelin synthesis \citep{Fads1_mouse,Sphingo_FattyLiver}. Less is known about ALMS1, although recent work suggests it contributes to nonalcoholic fatty liver disease \citep{ALMS1}.

\vspace{1mm}
\noindent \textbf{Factor 5: Hepatic lipid metabolism.} Factor 5's 19 significant SNPs map to either FADS1/2/3 or SLCO1B1. Unlike factors 4 and 7, factor 5 is characterize by a substantial number of outlier metabolites with similar large negative loadings (Figure~\ref{Figure:IndirectPathway}(a)), suggesting these metabolites have related functions and can be used to help infer factor 5's function. Remarkably, 24 out of the 25 outliers with loadings $<-0.4$ were omega-3 LC-PUFAs, which is congruent with the observation that the FADS polymorphisms we identify reduce the production of these anti-inflammatory LC-PUFAs in the liver \citep{FADS_PUFA}. Relatedly, SLCO1B1's polymorphisms are potential markers for nonalcoholic fatty liver disease \citep{SLCO_bile,SLCO_bile2}.


\subsection{Direct effects}
\label{subsection:RealData:Direct}
\begin{figure}
\centering
\includegraphics[width=0.8\textwidth]{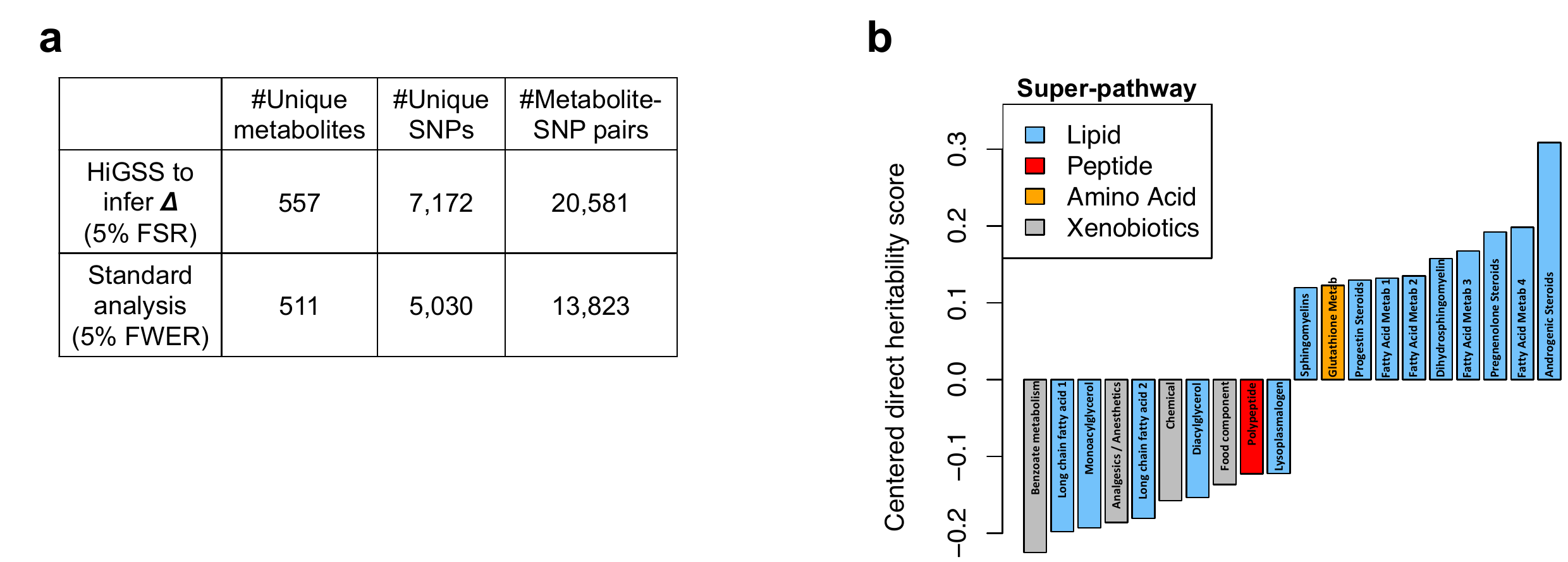}
\caption{Inference on $\bm{\Delta}$. (\textbf{a}): The FSR is the average local false sign rate. (\textbf{b}): Mean-centered direct heritability scores for the 10 least and most heritable sub-pathways.}\label{Figure:DirectEffects}
\end{figure}

We lastly used Algorithm~\ref{algorithm:HIGSS} and our pathway-guided prior defined in \ref{subsection:Direct} to infer $\bm{\Delta}$. Figure~\ref{Figure:DirectEffects}(a) contains an overview of our results. We first compare our results from HiGSS to those from a standard mtGWAS analysis, which uses the entries of $\hat{\bm{B}}$ and their standard errors to test $H_{0,sm}: \bm{B}_{sm}=\bm{G}_{s*}^{\top}\bm{L}_{m*} + \bm{\Delta}_{sm} = 0$ and subsequently controls the family-wise error rate with a Bonferroni correction. We identify 49\% more metabolite-SNP pairs than the standard analysis, which is quite remarkable given that we are only inferring direct effects whereas the standard analysis will reject pairs with significant direct or indirect effects.

While these results highlight the power of our method, interpreting them remains a challenge given the number of significant associations. We therefore considered performing a more interpretable analysis at the pathway level to determine the sub-pathways that have the largest direct effects. To do so, we used sub-pathway direct heritability scores defined in \eqref{equation:DHS}, where a large score indicates metabolites in that sub-pathway are more heritable. Figure~\ref{Figure:DirectEffects}(b) contains the results for the 10 least and most heritable sub-pathways. Unsurprising, xenobiotic sub-pathways, which typically contain exogenous metabolites, are the least heritable. The most heritable sub-pathway is ``Androgenic Steroids'', which is congruent with existing results from twin studies suggesting the heritability of circulating androgens may be as high as 50\% \citep{Androgen}.



\section{Conclusion}
\label{section:Conclusions}
In this work we developed a model, theoretical framework, and set of methods to perform GWAS of high dimensional phenotypes using summary statistics that explicitly model pleiotropy, beget fast computation, and facilitate the use of biologically informed priors. Given the importance of mtGWAS, we spent considerable effort developing hierarchical Dirichlet priors for indirect and direct genetic effects that foster interpretable inference at the metabolic pathway level.

Our real data analysis in Section~\ref{section:RealData} demonstrates the utility of our approach, where we were able to use our hierarchical Dirichlet process prior for indirect effects to couple metabolite sub-pathways by clustering them into biologically meaningful groups. We showed these were useful for interpreting factors, thereby providing putative mechanisms by which genotype impacts metabolite levels. Results using our method also suggested all observed metabolites were genetically regulated, which recapitulated results from a study with over 20 times as many samples as our data example. These conclusions were not possible using the standard analysis, which could only identify genetic associations for 67\% of metabolites and highlights the power of our approach.


\printbibliography

\newpage
\noindent{\LARGE \begin{center}Appendix for ``A statistical framework for GWAS of high dimensional phenotypes using summary statistics, with application to metabolite GWAS'' \end{center}}

\setcounter{section}{0}
\setcounter{Algorithm}{0}
\setcounter{figure}{0}
\renewcommand{\thesection}{A\arabic{section}}  
\renewcommand{\thesubsection}{A\arabic{subsection}} 
\renewcommand{\theAlgorithm}{A\arabic{Algorithm}}
\renewcommand{\thefigure}{A\arabic{figure}}

\section{The dBEMA algorithm}
\label{appendix:section:dBEMA}
Here we present our algorithm dBEMA, which we described in Section~\ref{subsection:K} and mirrors Algorithm 2 in \citet{ke2021estimation}. The data matrix input $\hat{\bm{B}}$ is as defined in \eqref{equation:Bhat}. To be consistent with \citet{ke2021estimation}, we parameterize the gamma distribution as $\text{Gamma}(\theta,\theta/\phi) \stackrel{\text{d}}{=} \phi \times \text{Gamma}(\theta,\theta)$.

Assume $S=S(N)$, $M=M(N)$, $S/N \to \gamma_S\in(0,\infty)$, and $M/N \to \gamma_M \in (0,\infty)$, and let $F_{R}$ and $F_{\sigma^2,N}$ be as defined in Section~\ref{subsection:K}. In what follows, we let $F_{\theta,\phi,N}$ be the empirical distribution of the non-zero eigenvalues of $N^{-1}\tilde{\bm{E}}^{\top}\tilde{\bm{E}}$ assuming the $M$ atoms of $F_{\sigma^2,N}$ were drawn from a $\text{Gamma}(\theta,\theta/\phi)$. We note that $F_{\theta,\phi,N}$ will converge to a limiting distribution whose Stieltjes transform is given in Lemma 1 of \citet{onatski2010determining} if we assume the atoms of $F_{\sigma^2,N}$ are drawn from a truncated gamma distribution. This is the same assumption made in \citet{ke2021estimation}. 


\begin{Algorithm}[dBEMA]
\label{appendix:algorithm:dBEMA}
\textit{Input data}: Standardized GWAS estimates $\hat{\bm{B}} \in \mathbb{R}^{S \times M}$, sample size $N$, parameters $\alpha,\beta \in (0,1)$, grid $0 < \theta_1 < \cdots < \theta_G$, and $B \in \mathbb{Z}_{>0}$.\\
\textit{Output}: An estimate for $K$, $\hat{K}$.
\begin{enumerate}
\item Define $\tilde{N} = \min(N,S,M)$ and $\hat{s}_1^2 \geq \cdots \geq \hat{s}^2_{\tilde{N}}$ to be the non-zero eigenvalues of $N^{-1}\hat{\bm{B}}^{\top} \hat{\bm{B}}$. Estimate $\theta$ and $\phi$ using the following steps:
\begin{enumerate}[label=(\roman*)]
    \item For each $k \in \{\lfloor (\alpha/2)\tilde{N} \rfloor, \ldots, \lfloor (1-\alpha/2)\tilde{N} \rfloor\}$ and $j \in [G]$, determine $q_{kj}$, an estimate of the $(k/\tilde{N})$th quantile of $F_{\theta_j,1,N}$.
    \item Let $\hat{\phi}_j =( \sum_k q_{kj}\hat{s}^2_k)/(\sum_k q_{kj}^2)$ for all $j \in [G]$.
    \item Define $j^* = \text{argmin}_{j \in [G]} \sum_k ( \hat{s}^2_k - \hat{\phi}_j q_{kj}  )^2$ and $\hat{\theta}= \theta_{j^*}$, $\hat{\phi} = \hat{\phi}_{j^*}$.
\end{enumerate}
\item Estimate the distribution of $N^{-1}\tilde{\bm{E}}^{\top}\tilde{\bm{E}}$'s maximum eigenvalue using the following steps:
\begin{enumerate}[label=(\roman*)]
    \item Let $\tilde{\bm{E}}^{(b)} \sim MN(0, \tilde{\bm{\Sigma}}_1,\tilde{\bm{\Sigma}}_2)$, where $\tilde{\bm{\Sigma}}_1\in\mathbb{R}^{S \times S}$ and $\tilde{\bm{\Sigma}}_2\in\mathbb{R}^{M \times M}$ are diagonal matrices whose elements are sampled from $F_R$ and $\text{Gamma}(\hat{\theta},\hat{\theta}/\hat{\phi})$, respectively.
    \item Let $\tilde{\lambda}_1^{(b)}$ be the largest eigenvalue of $N^{-1}\{\tilde{\bm{E}}^{(b)}\}^{\top}\tilde{\bm{E}}^{(b)}$.
    \item Repeat steps (i)-(ii) and for $b=1,\ldots,B$.
\end{enumerate}
\item Let $\hat{K}$ be the number of eigenvalues of $N^{-1}\hat{\bm{B}}^{\top}\hat{\bm{B}}$ that exceed the $1-\beta$ quantile of $\{\tilde{\lambda}_1^{(b)}\}_{b=1}^B$.
\end{enumerate}
\end{Algorithm}
In all simulations and our data application we let $\alpha=0.4$, $\beta = 0.1$, and $B=500$. Instead of specifying a grid $\theta_1,\ldots,\theta_G$, we estimated $\theta$ (and subsequently $\phi$) using a line search with lower and upper end points equal to 0.1 and 5. Lastly, we determined the quantiles $q_{kj}$ using a modified version of Algorithm 3 in \citet{ke2021estimation}, which consisted of running step 2 but replacing the $\text{Gamma}(\hat{\theta},\hat{\theta}/\hat{\phi})$ with $\text{Gamma}(\theta_j,\theta_j)$ and $\tilde{\lambda}_1^{(b)}$ with $\tilde{\lambda}_k^{(b)}$, the $k$th largest eigenvalue of $N^{-1}\{\tilde{\bm{E}}^{(b)}\}^{\top}\tilde{\bm{E}}^{(b)}$. We defined $q_{kj}=B^{-1}\sum_b \tilde{\lambda}_k^{(b)}$.

\section{Gibbs sampling using the Chinese restaurant franchise}
\label{appendix:section:GibbsSampler}

To perform inference on the parameters in model \eqref{equation:HDP}, we use a Gibbs sampler based on the Chinese restaurant franchise (CRF). We start with an analog of the CRF process. Each pair of parameters $(\mu,\phi^2)$ corresponds to a dish, a superpathway $p$ corresponds to a restaurant, a subpathway $b\in [B_p]$ corresponds to a customer. 
Here, each customer is a group of metabolites who usually show the same preference for a dish at a table except for a few outliers. 
We use $n_{pt}$ to denote the number of customers sitting at table $t$ in restaurant $p$ and $m_{pk}$ to denote the number of tables serving dish $k$ in restaurant $p$. Let $n_{p.}=\sum_t n_{pt}$, $m_{.k}=\sum_p m_{pk}$ and $m_{..}=\sum_k m_{.k}$. We use $t_{pb}$ to denote the table a subpathway $b$ of superpathway $p$ sits at. $k_{pt}$ denotes the dish index table $t$ in restaurant $p$ serves. 

Our base model described in Section \ref{subsection:Indirect} is
\begin{equation}\label{equ:base}
    H=\pi_{0,0}\delta_{(0,0)}+(1-\pi_{0,0})\Bprob(\mu)\times \textmd{dIG}(\alpha,\beta)
\end{equation}
where $\pi_{0,0}$ is the prior probability that a subpathway has loadings being exactly 0. $\Bprob(\mu)$ denotes the prior distribution of mean values, which is a mixture of spike and slab
\begin{equation*}
    \Bprob(\mu)=\pi_0\delta_0+\sum_{v=1}^V \pi_v N(\mu;0,\lambda_v^2)
\end{equation*}
where a nonzero $\pi_0$ allows for shrinking the means of subpathways to 0. The slab part is an approximate a Horseshoe prior, which is finite mixture of multiple normal distributions.
The standard deviations $\{\lambda_v\}$ were chosen as the 0.2, 0.4, 0.6, 0.8 and 0.9 quantiles of a half Cauchy distribution $C^+(0,1)$ and their corresponding weights $\{\pi_v\}$ are the standardized and scaled densities of $\{\lambda_v\}$,
\begin{equation*}
    \pi_v'=C^+(\lambda_v;0,1),\ \ \ \pi_v=\frac{(1-\pi_0)\pi_v'}{\sum \pi_v'}
\end{equation*} 
$\textmd{dIG}(\alpha,\beta)$ denotes a discritized inverse-gamma prior for $\phi^2$ with shape $\alpha$ and scale $\beta$,
\begin{equation*}
    \textmd{dIG}(\alpha,\beta)= \sum_{w=1}^W p_w \delta_{\sigma_w^2}
\end{equation*}
where $\{\sigma_w^2\}$ were generated by applying the quantile function of a inverse-gammma ($\alpha$, $\beta$) distribution to the values $\{1/51,2/51,...,50/51\}$ and their standardized densities were used as their weights $p_w$. 

For each factor $k=1,\ldots,K$, after the mean and variance $(\mu,\phi^2)$ of subpathway $b$ is sampled from the HDP model with base distribution $H$, concentration parameters $\gamma$ and $\alpha_0$, the corresponding entries of $\bm L$ are generated by
\begin{equation*}
    L_{pbm,k} \sim \pi_{out}U[a_u,b_u]+(1-\pi_{out})\mathcal{N}(\mu,\phi^2),\ m\in [M_{pb}]
\end{equation*}
The estimate $\hat{\bm L}$ of $\bm L$ is distributed as
\begin{equation*}
    \hat{\bm L}_{pbm,k}|\bm L_{pbm,k}\sim \mathcal{N}(L_{pbm,k},s_{mk}^2)
\end{equation*}
where $s_{mk}^2$ is the variance of the estimator $\hat{\bm L}_{pbm,k}$ given $\bm L_{pbm,k}$. 
Here we used a scaling of $\hat{\bm L}$ different from Algorithm \ref{algorithm:HIGSS}, which was $\hat{\bm L}=\sqrt{M}\bm V$ with $M$ being the number of metabolites and $\bm V$ being the first $K$ right singular vectors as defined in Algorithm \ref{algorithm:HIGSS}.
With the new scaling, each column of $\hat{\bm L}$ has norm $\sqrt{M}$. Scaling $\hat{\bm L}$ in this way avoids adjusting the hyperparameters in the base model according to the strength of each factor and we can rescale $\hat{\bm L}$ and the  parameters learned using Gibbs sampler afterwards. According to Theorem \ref{theorem:Lhat} \ref{theorem:Lhat:L}, with $\hat{\bm L}=\sqrt{M}\bm V$, we have
\begin{equation*}
    s_{mk}^2=\frac{M}{\Gamma_{kk}^2}(\frac{1}{N}+\frac{1}{S})\hat\sigma_m^2
\end{equation*}

We put Beta(1,1) priors on both $\pi_{0,0}$ and $\pi_0$, and a Beta(1/5,999/5) prior on $\pi_{out}$. The latter reflects our prior belief that we will have on average 1 outlier out of 1,000 metabolites and no more than 5 outliers with probability 0.95. Following \citep{HDP}, we put Gamma(1,1) priors on the concentrations parameters $\gamma$ and $\alpha_0$. But different from \citep{HDP}, we discretrized the Gamma prior for $\gamma$ and $\alpha_0$ to avoid time-consuming auxiliary variable sampling.

In our implemention, we first sample the table index $t_{pb}$ for the non-outlier metabolites of each subpathway, then update the dish indices $\{k_{pt}\}$ for tables and finally update the parameters $\{(\mu,\phi^2)\}$. The update of other relevant parameters will be demonstrated in the process. For simplicity of notations, the factor index $k$ is omitted because the Gibbs update will be applied to each column of $\hat{\bm L}$ independently and in parallel. The index $k$ will be used to indicate parameter $\theta_k$ in the content below. We use $\theta=(\mu,\phi^2)$ to denote the pair of parameters sampled from the HDP model (\ref{equation:HDP}). Thus, for subpathway $b$, the parameter it eventually consumes is $\theta_{kt_{pb}}=(\mu_{kt_{pb}},\phi^2_{kt_{pb}})$.

\textit{\textbf{Updating  outliers}.}
For each sub-pathway $b$, we first update the outliers, because outliers not generated from the HDP model and only the non-outlier entries will be included in the next round of Gibbs update. The posterior probability that a metabolite $m$ of subpathway $b$ is an outlier, given $\hat{\bm L}$ and other parameters, can be obtained by integrating $\bm L$ out,
\begin{equation*}
    \Bprob(\textmd{$m$ is an outlier} | \cdot)\propto \frac{\pi_{out}}{b_{out}-a_{out}}\left[ \Phi(\frac{b_{out}-\hat{\bm L}_{pbm}}{s_m})-\Phi(\frac{a_{out}-\hat{\bm L}_{pbm}}{s_m}) \right]
\end{equation*}
where $\Phi$ is the CDF of a standard normal distribution. The parameters of the uniform distribution, $a_{out}$ and $b_{out}$ should cover the range of $\hat{\bm L}_k$. In our implement, $a_{out}$ and $b_{out}$ were set to centered at zero and cover 1.25 times the range of $\hat{\bm L}_k$. 

The posterior probability that a metabolite is generated from the HDP model depends on the current Gibbs sample for sub-pathway $b$. If $b$ contains other non-outlier metabolites, which have been assigned to a table $t$ in the previous round of Gibbs update, then 
\begin{equation*}
    \Bprob(\textmd{$m$ is not an outlier}|\cdot)\propto 
    (1-\pi_{out})\Bprob(\hat{\bm L}_{pbm}|\theta_{k_{t_{pb}}},\sigma_m^2)
\end{equation*}
where
\begin{equation*}
   \Bprob(\hat{\bm L}_{pbm}|\theta_{k_{t_{pb}}},s_m)= N(\hat{\bm L}_{pbm};\mu_{k_{t_{pb}}},\phi^2_{k_{t_{pb}}}+s_m^2).
\end{equation*}
Otherwise, if currently there are no non-outlier metabolites in $b$ except for $m$, which could happen if $b$ contains only one metabolite or all other metabolites in $b$ are outliers, we need to integrate $\theta_{k_{t_{pb}}}$ out conditional on the current table and dish assignments,
\begin{equation*}
   \begin{split}
        \Bprob(\textmd{$m$ is not an outlier}|\cdot)\propto &
    (1-\pi_{out})\left[\sum_{t}\frac{n_{pt}}{n_p-1+\alpha_0}\Bprob(\hat{\bm L}_{pbm}|\theta_{k_{pt}},s_m)\right.\\
    + \frac{\alpha_0}{n_p-1+\alpha_0} &\left.\left(\sum_k\frac{m_{.k}}{m_{..}+\gamma}
    \Bprob(\hat{\bm L}_{pbm}|\theta_{k},s_m)+\frac{\gamma}{m_{..}+\gamma}\Bprob(\hat{\bm L}_{pbm}|H,s_m)\right) \right]
   \end{split}
\end{equation*}
where
\begin{equation*}\label{equ:nonconj}
    \Bprob(\hat{\bm L}_{pbm}|H,s_m)=\pi_{0,0}N(\hat{\bm L}_{pbm};0,s_m^2)+(1-\pi_{0,0})\iint N(\hat{\bm L}_{pbm};\mu,\phi^2+s_m^2)\Bprob(\mu)\Bprob(\phi^2) d\mu d\phi^2
\end{equation*}
The integral on the right hand side of the above equation can be easily calculated because we discretized $ \Bprob(\mu)$ and $\Bprob(\phi^2)$. Otherwise, the integral is difficult to evaluate because $N(\hat{\bm L}_{pbm};\mu,\phi^2+s_m^2)$ is not conjugate to the inverse-gamma prior due to the existence of $s_m^2$.
This discretized prior on $\phi^2$ helps overcome the non-conjugacy problem and avoids time-consuming Monte Carlo simulations.

\textit{\textbf{Sampling t}.}
We can treat $t_{pb}$ as the last variable being sampled in restaurant $p$ due to exchangeability. The likelihood of $t_{pb}=t$ given other parameters is proportional to the number of subpathways sitting at $t$ if $t$ is not empty, or proportional to $\alpha_0$ if $t$ is new. The likelihood of $\hat{\bm L}_{pb}$ given $t_{pb}$ is a product of the likelihoods of all non-outlier metabolites of subpathway.
 The likelihood of $\hat{\bm L}_{jg}$ given $t=t_{new}$ can be calculated by integrating out $k_{pt}$,
\begin{equation}
\Bprob(t_{pb}=t|\cdot)\propto 
 \begin{cases}
       n_{pt}^{-b}\prod_{m\in [M_{pb}]} \Bprob(\hat{L}_{pbm}|
\theta_{k_{pt}},s_m)& \textmd{if $t$ previously used}  \\
         \alpha_0\prod_{m\in [M_{pb}]}\Bprob(\hat{L}_{pbm}|t=t_{new},\bm\theta,\bm k,s_m)  & \textmd{if $t=t_{new}$}
    \end{cases} 
\end{equation}
and 
\begin{equation}
    \Bprob(\hat{L}_{pbm}|t=t_{new},\bm\theta,\bm k,s_m^2)=\sum_k\frac{m_{.k}}{m_{..}+\gamma}
    \prod_{m\in [M_{pb}]} \Bprob(\hat{\bm L}_{pbm}|\theta_{k},s_m)+\frac{\gamma}{m_{..}+\gamma}\prod_{m\in [M_{pb}]} \Bprob(\hat{\bm L}_{pbm}|H,s_m)
\end{equation}
If subpathway $g$ sits at $t_{new}$, we need to assign a dish to $t_{new}$ by sampling $k_{pt_{new}}$ conditional on data and this will be introduced in the next step. As a result of updating $t_{pb}$, some table may become unoccupied, then with probability 0 they will be sampled again in later iterations because $n_{pt}=0$. As a result, we remove unoccupied tables after updating $t_{pb}$ each time. Since removing tables, as well as updating $k_{pt}$ in the next step, can potentially cause some dishes unoccupied, we will also delete unoccupied dishes after updating $\bm t$ and $\bm k$.

\textit{\textbf{Sampling k}.}
The procedure sampling $k_{pt}$ is similar to sampling $t_{pb}$. First we assume the exchangeability of dish assignments and treat $k_{pt}$ as the last variable to be sampled. Since a table can serve multiple subpathways, the likelihood of $k_{pt}=k$ will contain production across subpathways and metabolites in those subpathways,
\begin{equation}
    \Bprob(k_{pt}=k|\cdot )
       \propto  \begin{cases}
           m_{.k} \prod_{b:t_{pb}=t}\prod_{m\in [M_{pb}]} \Bprob(\hat{\bm L}_{pbm}|\theta_k,s_m) &\text{if $k$ is previously used} \\
           \gamma \prod_{g:t_{pb}=t}\prod_{m\in [M_{pb}]} \Bprob(\hat{\bm L}_{pbm}|H,s_m) &\text{if $k=k_{new}$}
       \end{cases}
\end{equation}
If $k_{t}=k_{new}$, we will draw a new pair of parameters $\theta_{k_{new}}=(\mu_{new},\sigma_{new})$ from $H$ conditional on  $\hat{\bm L}_{pbm}$ and $s_m$ of all subpathways sitting at table $t$, i.e. $t_{pb}=t$. We will discretize the inverse-gamma prior on $\phi^2$ because of the non-conjugacy problem arising from $s_m^2$. See next step for details.

\textit{\textbf{Updating concentration parameters}.}
Updating $\gamma$ and $\alpha_0$ do not depend on the values of $\bm\theta$, it depends on only the number of dishes and tables, which is fixed after sampling $t$ and $k$. Thus, it was implemented before updating $\bm\theta$.

For updating of concentration parameters, we used the likelihood of $\gamma$ and $\alpha_0$ in \citep{HDP} conditional on $K$, the total number of dishes, $\{T_p\}_1^P$, the number of tables in superpathway $p$ and $\{B_p\}_1^P$, the number of subpathways in superpathway $p$. 
In our generative model (\ref{equation:HDP}) for $\bm L$, the concentration parameter $\alpha_0$ is shared across all superpathways. Thus, we have
\begin{equation*}
    \Bprob(T_1,...T_P|\alpha_0,B_1,...,B_P)=\prod_{p=1}^P s(B_p,T_p)\alpha_0^{T_p}\frac{\Gamma(\alpha_0)}{\Gamma(\alpha_0+B_p)}
\end{equation*}
and
\begin{equation*}
    \Bprob(K|\gamma, \sum_{p}T_p)=s(\sum_{p}T_p,K)\gamma^K\frac{\Gamma(\gamma)}{\Gamma(\gamma+\sum_{p}T_p)}
\end{equation*}
where $s(n,m)$ is unsigned Stirling numbers of the first kind.
Instead of using auxiliary variable sampling or adaptive rejection sampling to iteratively sample $\gamma$ and $\alpha_0$, we discretized the gamma priors for $\gamma$ and $\alpha_0$ and update their values by sampling from their grids using their posterior probabilities.
The grids for both $\gamma$ and $\alpha_0$ were set as $\{0.1,0.2,...,0.9,1,2,...,10\}$.

\textit{\textbf{Updating $\bm\theta$}.}
Not only sampling new parameters, but also updating old parameters require sampling the mean values and variances shared by subpathways conditional on metabolite loading $\hat{\bm L}_{pbm}$ and the variances of error $s_m^2$. Since parameters $\bm\theta$ are shared by all restaurants(superpathways), we will update $\theta_k$ conditional on all non-outlier metabolites consuming dish $k$ across superpathways and subpahtways. Let $\{k\}$ denote all metabolites consuming dish $k$, we have
\begin{equation}
    \Bprob\left( \theta_k=(0,0)|\cdot\right)\propto
    \pi_{0,0}\prod_{m\in \{k\}}N(\hat{\bm L}_{pbm};0,s_m^2)
\end{equation}
and
\begin{equation}
     \Bprob\left( \theta_k\neq(0,0)|\cdot \right)\propto
     (1-\pi_{0,0})\iint \prod_{m\in \{k\}}N(\hat{\bm L}_{m};\mu,\phi^2+s_m^2) d\mathcal{P}(\mu) d\textmd{IG}(\phi^2)
\end{equation}
If $\theta_k\neq(0,0)$, we will draw indicator variables 
\begin{equation}
    \Bprob\left( Z_{v,w}=1|\cdot\right)\propto p_vp_w \prod_{m\in \{k\}}N(\hat{\bm L}_{m};0,\lambda_v^2+\sigma_w^2+s_m^2)
\end{equation}
If $Z_{0,r}=1$, we let $\theta_k=(0,\sigma_w)$ directly. Otherwise, we  draw the value of $\mu_k$ given $\lambda_v$ and $Z_{v,w}$,
\begin{equation}
    \sigma_k=\sigma_w,\ \ \mu_k\sim N\left(\frac{\sum_{m\in \{k\}}\frac{\hat{\bm L}_m}{\sigma_w^2+s_m^2}+\frac{0}{\lambda_v^2}}{\sum_{m\in \{k\}}\frac{1}{\sigma_w^2+s_m^2}+\frac{1}{\lambda_v^2}},\frac{1}{\sum_{m\in \{k\}}\frac{1}{\sigma_w^2+s_m^2}+\frac{1}{\lambda_v^2}}\right)
\end{equation}

\textit{\textbf{Updating $\pi_{out},\pi_{0,0},\pi_0$}.}
The updates of $\pi_{out}$, $\pi_{0,0}$ and $\pi_0$ can be obtained by
\begin{equation*}
    \pi_{out}\sim\textmd{Beta}(\frac{1}{5}+\#outliers,\frac{999}{5}+M-\#outliers)
\end{equation*}
\begin{equation*}
    \pi_{0,0}\sim\textmd{Beta}(1+\#spikes,1+K-\#spikes)
\end{equation*}
\begin{equation*}
    \pi_{0}\sim\textmd{Beta}(1+\#centers,1+K-\#spikes-\#centers)
\end{equation*}
where $\#outliers$, $\#spikes$ and $\#centers$ denotes the number of outliers, the number of parameters $\theta=(0,0)$ and the number of parameters with mean 0 but non-zero variances. $M$ and $K$ denote the total number of metabolites and number of parameters in the current round of Gibbs update.

\section{Estimating direct effects in mtGWAS: Setting grid points}
\label{appendix:section:DirectEffects}
Here we describe our procedure to define the grid points in $\mathcal{X}^{(\pi)}$ and $\mathcal{X}^{(\varphi^2)}$ defined in Section~\ref{subsection:Direct}. Briefly, we first estimate $\{\pi_{pbm}\}$ and $\varphi^2$ via maximum likelihood as
\begin{align*}
    \{ \{\hat{\pi}_{pbm}\}, \hat{\varphi}^2 \} = \mathop{\text{argmax}}_{\{\pi_{pbm}\},\varphi^2 } \Bprob( \hat{\bm{\Delta}} \mid \{\pi_{pbm}\},\varphi^2 ),
\end{align*}
assuming
\begin{align*}
    &\hat{\bm{\Delta}}_{s,pbm} \mid \bm{\Delta}_{s,pbm} \sim N(\bm{D}_{ss}^{1/2}\bm{\Delta}_{s,pbm}, \hat{\sigma}_{pbm}^2)\\
    &\bm{\Delta}_{s,pbm} \mid \pi_{pbm},\varphi^2 \sim (1-\pi_{pbm})\delta_0 + \pi_{pbm} N(0,\varphi^2 \hat{\sigma}_{pbm}^2/\bm{D}_{ss}).
\end{align*}
We then define $\mathcal{X}^{(\pi)}$ and $\mathcal{X}^{(\varphi^2)}$ to each be sets with 100 elements, where $\mathcal{X}^{(\varphi^2)} = \{\hat{\varphi}^2/2,\ldots,2\hat{\varphi}^2\}$ contains equally spaced elements. We set $\mathcal{X}^{(\pi)}$ so that its minimum element is $10/S$, its maximum element is 1.5 times the 0.99 quantile of $\{\hat{\pi}_{pbm}\}$, and is equally spaced on a log scale.

\section{Simulating \textbf{L} for experiments}
\label{appendix:section:SimL}
To simulate a $\bm L$ of $K$ factors, a super-sub pathway structure of metabolites need to be specify first. In our implement the number of subpathways of a superpathway and sizes of subpathways were randomly sampled from Poisson distributions. In summary, we generated a metabolite archives containing 3 superpathways and 257 metabolites.

Given the pathway structure of metabolites, each column of $\bm L$ was generated independently. The generating procedure of $\bm L$ exactly followed the Chinese restaurant franchise construction of the HDP model in Equation \eqref{equ:base}. The global shrinkage parameter $\tau$ of the Horseshoe prior was set as 0.2, outlier rate $\pi_{out}$ was 0.01, spike rate $\pi_{0,0}$ was 0.1, $\pi_0$ was randomly sampled from a Beta(2,1) distribution and the scale and rate for the inverse gamma distribution where $\phi^2$ will be drawn were set as 2 and 1. To make sure the outliers were recoverable, the outlier entries were generated beyond 2 standard deviations away from the mean of the subpathway, but they cannot exceed (-8,8). The simulation procedure allowed us to track which subpathways were spikes, which metabolites were outliers and which subpathways shared the same parameters. Finally, all entries of $\bm L$ were multiplied by 0.4 to make the strength of signals comparable to the noise level.

\begin{figure}
\centering
\includegraphics[width=1\textwidth]{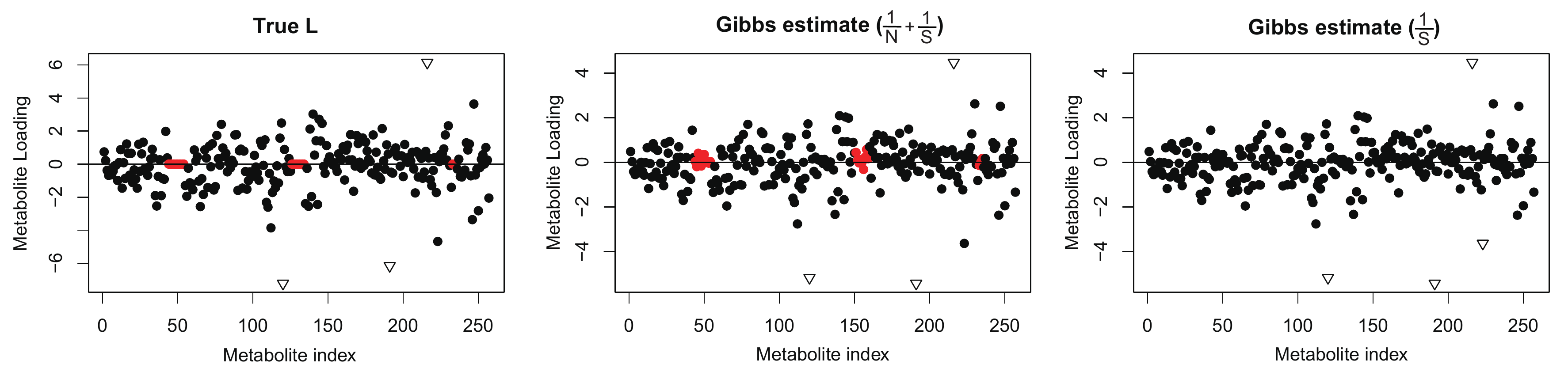}
\caption{\textit{Left}: True $\bm L$ simulated using the model in Equation \ref{equation:HDP}. Spike pathways are marked in red. Outliers are marked by inverted triangles. \textit{Middle}: Estimates for spikes and outliers with the correct asymptotic variance of $\hat{\bm{L}}$. Estimated spike pathways are marked in red. Estimated Outliers are marked by inverted triangles. \textit{Right}: Estimates for spikes and outliers with the incorrect variance for $\hat{\bm{L}}$ derived by ignoring the correlation between SNPs. Estimated spike pathways are marked in red. Estimated Outliers are marked by inverted triangles.}\label{figure:simuL_spikes}
\end{figure}

We lastly show that the Gibbs sampler described in Section \ref{appendix:section:GibbsSampler} can identify spikes and outliers using the asymptotic variance of $\hat{\bm{L}}$ derived in Theorem \ref{theorem:Lhat}. 
To do so, we simulated a phenotype matrix $\bm Y$ with 257 metabolites, 5000 SNPs and sample size being 500, making the ratio $\frac{S}{N}=10$. $\bm Y$ contained $K=1$ factor to simplify the simulation by avoiding identifiability issues that arise with more than one factor.
The $\bm L$ used to simulate $\bm Y$ is shown in Figure \ref{figure:simuL_spikes} \textit{Left}, which contains 3 spike subpathways (marked in red) and 3 outliers (marked by 
inverted triangles). The Gibbs sampler described in Section \ref{appendix:section:GibbsSampler} was applied to the estimate $\hat{\bm L}$ twice, once assuming $\V(\hat{\bm L}_{mk}\mid \bm L) = (1/N+1/S)\sigma_m^2$ and another assuming $\V(\hat{\bm L}_{mk}\mid \bm L) = (1/S)\sigma_m^2$. The second is the variance one would assume if they treat $\bm{R}$ in \eqref{equation:Bhat} as the identity \citep{BCconf}. The results are shown in Figure \ref{figure:simuL_spikes} \textit{Middle} and \textit{Right}. A subpathway was concluded as a spike if the posterior probability that it consumed (0,0) as its parameter exceeded 0.95. Outliers were defined as the metabolites whose posterior probability of being an outlier was greater than 0.95. It can be observed that the Gibbs sampler was able to identify spikes and outliers correctly with the right variance, but it failed identifying any spike subpathways and overestimates the number of outliers.

\end{document}